\documentclass{jpp}

\usepackage[utf8]{inputenc}
\usepackage[T1]{fontenc}
\usepackage{amsmath}

\usepackage{graphicx}
\graphicspath{ {./images/} }

\usepackage{authblk}
\usepackage{subcaption}
\usepackage{tikz,siunitx,mwe}
\usepackage[thinc]{esdiff}
\usepackage{float}
\usepackage{gensymb}
\usepackage{amssymb}
\usepackage{mathrsfs}
\shorttitle{Intermittency of density fluctuations and zonal-flow generation}
\shortauthor{A. Sladkomedova et al.}

\title{Intermittency of density fluctuations and zonal-flow generation in MAST edge plasmas}

\author{A. Sladkomedova\aff{1} \footnote{Present address: Tokamak Energy, Ltd., 173 Brook Dr, Milton, Abingdon OX14 4S, UK}
  \corresp{\email{alsu.sladkomedova@tokamakenergy.co.uk}},
  I. Cziegler\aff{1},
  A. R. Field\aff{2},
  A. A. Schekochihin\aff{3,4},
  D. Dunai\aff{5},
  P. G. Ivanov\aff{2,3}
 and the MAST-U Team and the EUROfusion MST1 Team}

\affiliation{\aff{1}York Plasma Institute, Department of Physics, University of York, York, YO10 5DQ, UK
\aff{2}UKAEA/CCFE, Culham Science Centre, Abingdon, Oxon, OX14 3DB, UK
\aff{3}Rudolf Peierls Centre for Theoretical Physics, University of Oxford, Oxford, OX13PU, UK
\aff{4}Merton College, Oxford, OX145D, UK
\aff{5}HUN-REN Centre for Energy Research, Budapest, 1121, Hungary}

\date{}
\begin{document}

\maketitle

\begin{abstract}

The properties of the edge ion-scale turbulence are studied using the beam emission spectroscopy (BES) diagnostic on MAST. Evidence of the formation of large-scale high-amplitude coherent structures, filamentary density blobs and holes, 2$-$4 cm inside the plasma separatrix is presented. Measurements of radial velocity and skewness of the density fluctuations indicate that density holes propagate radially inwards, with the skewness profile peaking at 7$-$10 cm inside the separatrix. Poloidal velocities of the density fluctuations measured using cross-correlation time delay estimation (CCTDE) are found to exhibit an intermittent behaviour. Zonal-flow analysis reveals the presence of poloidally symmetric coherent oscillations $-$ low-frequency (LF) zonal flows and geodesic acoustic modes (GAM). Shearing rates of the observed zonal flows are found to be comparable to the turbulence decorrelation rate. The observed bursts in density-fluctuation power are followed by quiescent periods with a transient increase in the power of sheared flows. Three-wave interactions between broadband turbulence and a GAM are illustrated using the autobispectral technique. It is shown that the zonal flows and the density-fluctuation field are nonlinearly coupled and LF zonal flows mediate the energy transfer from high- to low-frequency density fluctuations.

\end{abstract}

\section{Introduction}
\label{intro}

Edge-plasma fluctuations represent a complex system exhibiting intermittent behaviour due to the presence of coherent structures $-$ long-lived plasma filaments with large-amplitude positive (blobs) and negative (holes) density fluctuations. The link between the pedestal pressure and confinement of the core plasma makes it crucial to understand edge-plasma fluctuations. The power-exhaust problem in future fusion reactors requires prediction of the heat loads on the plasma-facing components and hence of the dynamics of the edge plasma. Edge fluctuations inside the separatrix might influence the width of the scrape-off-layer (SOL) through the turbulence spreading \citep{Manz2015, Grenfell2019, Silvagni2020}, suggesting non-local origin of transport in the SOL. Another important implication of the presence of coherent structures in the plasma edge is that radial propagation of density holes may represent a mechanism for impurity transport in the edge plasma \citep{Kotschenreuther2004, Krasheninnikov2008}.

Analytical and numerical works have provided evidence for the existence of coherent structures in the bad-curvature region of tokamaks \citep{Krasheninnikov2001,Dippolito2002,Bian2003}, with some of the studies highlighting the formation of blob$-$hole pairs \citep{Krasheninnikov2008,Aydemir2005,Kendl2015,Churchill2017}. The formation mechanism of such structures is still debated; they are likely a feature of nonlinear turbulence saturation. Thus, zonal flows near their birth location are believed to play an important role in the ejection process of blobs \citep{DIppolito2011}. It has been shown using simulations of curvature-driven ITG turbulence that long-lived coherent structures are generated within the region of strong flow shear, with increase in the zonal-shear amplitude anticorrelating with the bursts in the turbulent heat flux \citep{Ivanov2020}. This and other theoretical works \citep{DIppolito2011,Garcia2005} suggest that coherent structures are part of a self-organised system regulated by sheared flows. Coherent structures have also been observed in gyrokinetic simulations in the presence of a strong equilibrium $E \times B$ flow shear \citep{VanWyk2016,VanWyk2017}.

Early experiments focused on edge turbulence and observation of coherent structures include studies on the Caltech tokamak using a 2-D probe array \citep{Zweben1985} and on TFTR \citep{Zweben1989} using a fast video camera. Later experiments presented evidence of filaments on ASDEX Upgrade \citep{Benkadda1994} and using 2-D fast probes and beam emission spectroscopy (BES) on DIII-D \citep{Boedo2001}. Probes and the gas puff imaging diagnostic were used to investigate the intermittent transport on Alcator C-Mod \citep{Terry2005}. On MAST, blob filaments have been previously studied in the SOL and just inside the separatrix using fast cameras and Langmuir probes \citep{Kirk2006,Hnat2008,Ayed2009,Militello2012}.

The majority of experimental observations reported in the literature is devoted to the detection of blobs. Some of the works, e.g., on NSTX \citep{Maqueda2011}, ASDEX Upgrade \citep{Nold2010}, DIII-D \citep{Boedo2003}, and JET \citep{Xu2009}, feature the presence of both positive and negative density bursts existing on the opposite sides of their formation zones. The existence of a shear layer  coinciding with the region of formation of the blobs and holes has been observed just inside the separatrix on JET using a fast-reciprocating Langmuir probe array \citep{Xu2009}. It has been argued that blobs transfer energy to zonal flows via nonlinear interactions that lead to saturation of edge turbulence and suppression of mesoscale structures \citep{Xu2009}.

Zonal flows are toroidally symmetric plasma flows in the direction perpendicular to the magnetic field. They are nonlinearly driven by interactions between unstable modes and have a finite radial extent \citep{Diamond2005}. These spatially localised flows occur in two varieties $-$ stationary zonal flows \citep{Hinton1999}, which exist at low frequencies, and oscillating zonal flows, or the geodesic acoustic modes (GAM) \citep{Winsor1968}, which have a frequency that scales as $c_s/R$ with the sound speed $c_s$ and major radius $R$. Zonal flows are believed to play an important role in the regulation of plasma oscillations and hence plasma transport. There has been increasing evidence for the presence of zonal flows in tokamak plasmas, highlighting frequency-resolved nonlinear interactions between sheared flows and turbulence \citep{McKee2003, Holland2007, Nagashima2006, Liu2010, Xu2012, Hillesheim2012, Cziegler2015, Melnikov2017}, however, such measurements are still rare for spherical tokamaks \citep{Bulanin2016}.

This work studies ion-scale density fluctuations and zonal flows in the neutral-beam-heated L-mode plasmas in the MAST spherical tokamak using the BES diagnostic. We present evidence of high-amplitude coherent structures, which are blobs and density holes, existing amongst ambient turbulent fluctuations. By using statistical analysis and a velocimetry technique, we explore the characteristics of intermittent density fluctuations at $r/a=0.8-1.1$. The paper focuses, in particular, on the properties and impact of holes deep in the confined region. Our findings complement the existing picture of blob-dominated transport near the separatrix and in the SOL. The self-regulation of the system consisting of turbulence and longer-lived structures is addressed by analysing the dynamics of perpendicular flows. We find that the generation of GAM follows bursts in the density-fluctuation power originating from turbulence and radially moving holes. Bispectral techniques are used to illustrate nonlinear interactions between zonal flows and density oscillations.

This paper is organised as follows. Section \ref{Experim_setup} describes the experimental set-up, plasma parameters, and diagnostic tools. In section \ref{edfe_fl}, we explore edge density fluctuations and their radial dynamics. Statistical properties of the plasma-density fluctuations and coherent structures are discussed in section \ref{stat_fl} and section \ref{Coherentstructures}, respectively. Relation of statistical properties of density fluctuations to the equilibrium profiles is discussed in section \ref{Equilibrium}; radial propagation of the density structures is discussed in section \ref{Dynamics}. Section \ref{Zonal flows} details the intermittent character of the perpendicular velocity fluctuations (section \ref{v_vs_time}) and the identification of zonal flows (section \ref{mode}); the impact of the zonal flows on density fluctuations and their nonlinear interactions are discussed in sections \ref{Shear_rate} and \ref{nonlincoupl}, respectively. A summary is given in section \ref{summary}, followed by conclusions in section \ref{conc}.

\section{Experimental set-up}
\label{Experim_setup}
The experiments were performed on the spherical MAST tokamak \citep{Chapman2015}, with major radius $R=0.73$ m and aspect ratio $R/a=1.4$, where $a$ is the minor radius. The plasma parameters of the analysed discharges were: the plasma current during the flat-top phase $I_p=0.7$ MA, the toroidal magnetic field $B_0$=0.6 T, and the power of the deuterium neutral beam $P_{NBI}=2$ MW. Our analysis was carried out for L-mode deuterium plasmas during the current ramp-up and flat-top phases in the limiter and divertor plasma configurations.
Measurements of the ion-scale turbulence were  performed using the BES diagnostic. The diagnostic is based on the detection of the Doppler-shifted D-$\alpha$ line radiation of an injected neutral beam. The BES diagnostic on MAST consisted of 4$\times$8 channels in poloidal and radial directions, respectively, and thus provided a 2-D image of the density-fluctuation field. The diagnostic was capable of measuring the density fluctuations with radial and poloidal wavenumbers $k_{r,\theta} \rho_i \lesssim 1 $ at 2 MHz sampling frequency. More details on the diagnostic performance can be found from \citet{Field2014}.

The analysis was performed for the shots $\# 27292 - \#27298$ and $\# 27308 - \#27310$. The corresponding time periods were $0.109-0.143$ s, $0.109-0.145$ s, $0.109-0.165$ s, $0.109-0.146$ s, $0.109-0.150$ s, $0.109-0.164$ s, $0.109-0.164$ s, and $0.100-0.150$ s, $0.138-0.162$ s, $0.100-0.158$ s. The choice of the shots and the time ranges was dictated by the necessity of a sufficiently high signal-to-noise ratio. For the cases considered here, the measured fluctuation level exceeded the photon noise by more than a factor of 2.5 across at least five radial locations. The turbulence data suitable for our study were limited by the presence of strong fast-ion-driven magnetohydrodynamic (MHD) activity. Periods containing spurious signals from such MHD bursts were excluded from our analysis. The present study focuses on edge fluctuations at radial locations covering the range $r/a=0.8-1.1$, where $r$ is the minor radius of the magnetic flux surface, $a$ is the plasma minor radius at the boundary, and the corresponding distance to the separatrix is $\Delta R =R-R_{sep}= -15$ to $7$ cm, where $R$ is the major radius of the measurement location and $R_{sep}$ is the major radius of the separatrix at the outer mid-plane. 

\section{Edge fluctuations}
\label{edfe_fl}
\subsection{Statistical properties of edge fluctuations}
\label{stat_fl}

\subsubsection{Intermittency of density fluctuations}
\begin{figure}
    \centering
    \begin{subfigure}[b]{0.50\textwidth}
    	\includegraphics[trim={0 0 0 0},clip,width=\textwidth]{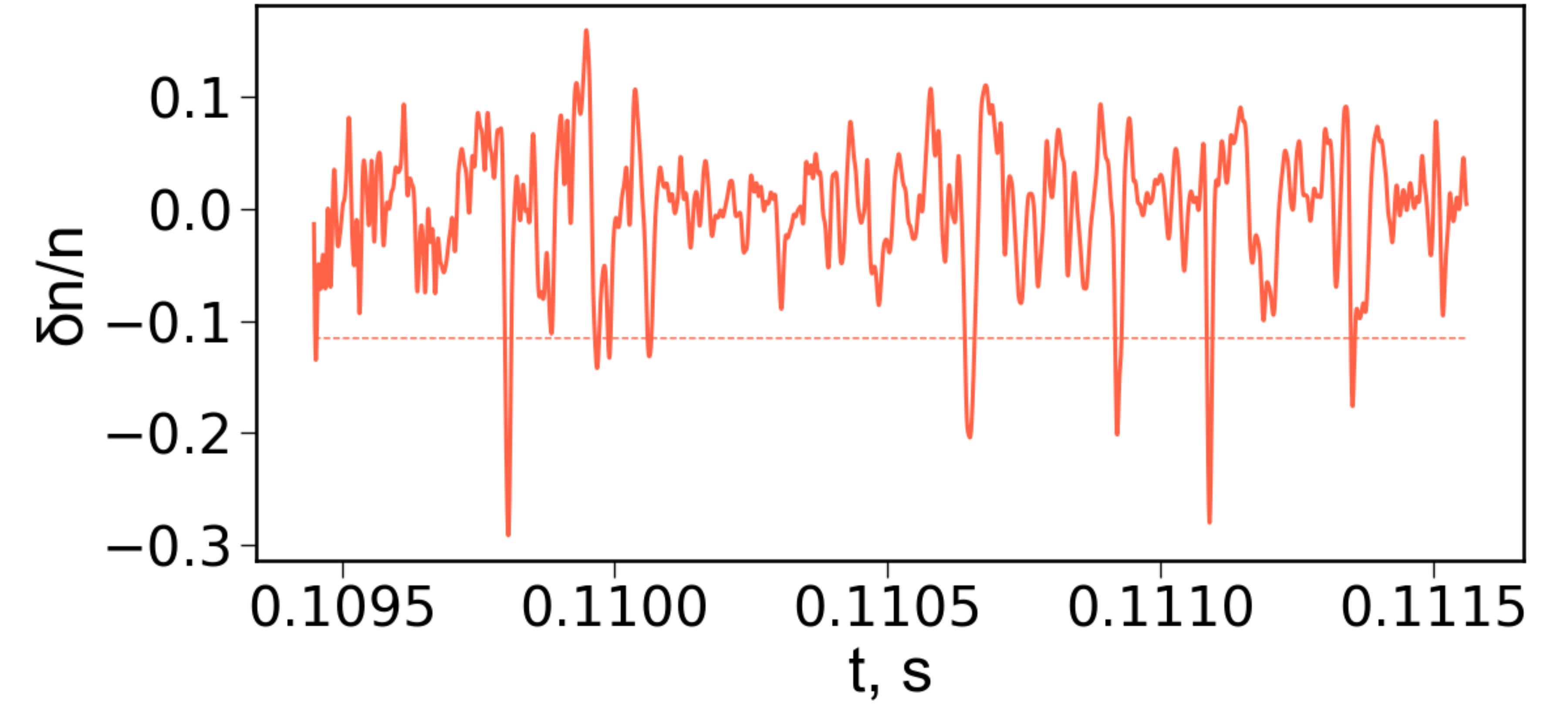}
	    \caption{}
    \end{subfigure}
    \hfill
    \begin{subfigure}[b]{0.52\textwidth}
    	\includegraphics[trim={0 0 0 0},clip,width=\textwidth]{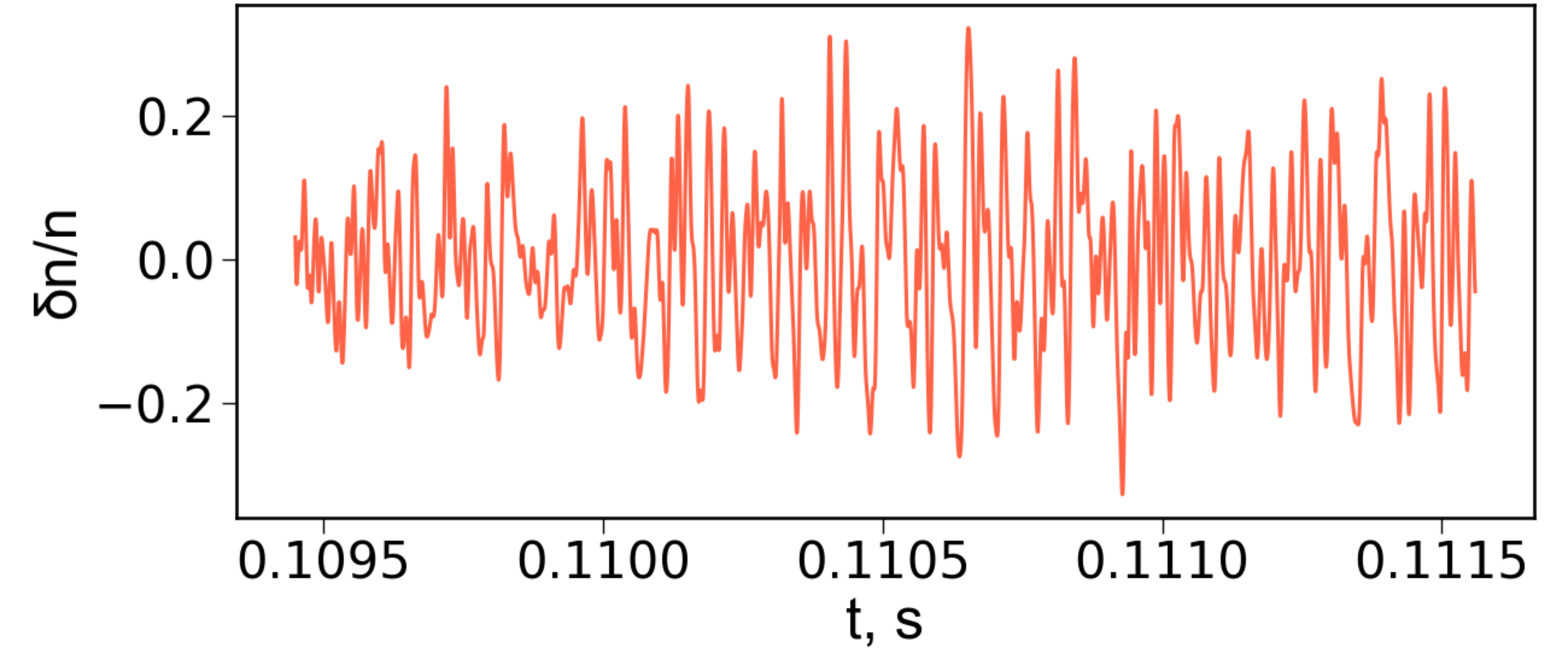}
    	\caption{}
    \end{subfigure}
    \hfill
	\begin{subfigure}[b]{0.52\textwidth}
    	\includegraphics[trim={0 0 0 0},clip,width=\textwidth]{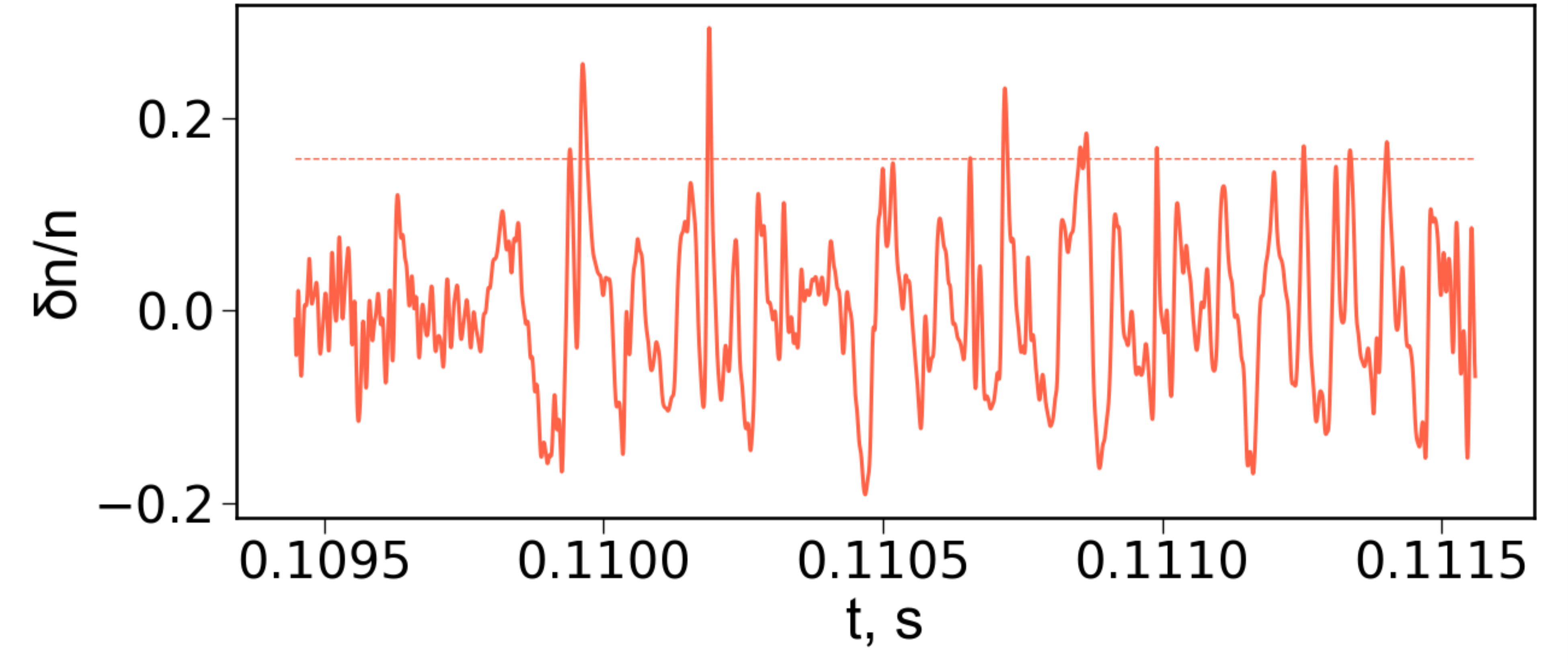}
    	\caption{}
    \end{subfigure}
    \caption{Time traces of relative density fluctuations at different radial locations in the shot $\#$27292: (a) $R=1.21$ m, $\Delta R=-9$ cm, $r/a=0.86$; (b) $R=1.26$ m, $\Delta R=-4$ cm, $r/a=0.88$; (c) $R=1.28$ m, $\Delta R=-2$ cm, $r/a=0.97$. The vertical coordinate of the measurement location was $Z=-0.01$ m.}
	\label{fig: timeseries}
\end{figure}

The time series of the relative density fluctuations at three different radial locations within the confined region are presented in Figure \ref{fig: timeseries}. The intermittent character of the fluctuations is evident but will presently be analysed quantitatively using higher-order moments of the fluctuation amplitude. The third moment, the skewness $S$, is a measure of the asymmetry of the probability distribution function (PDF) of the amplitude. The fourth moment, the kurtosis $K$, characterises the heaviness of the tails of the PDF.  These quantities are defined as
\begin{equation}
\label{skewdef}
 S=\frac{\mu _3}{\sigma^3}, \quad \mu _3=\frac{1}{M} \sum_{i=1}^M \left( \frac{\delta n}{n}-\Biggl \langle \frac{\delta n}{n} \Biggr \rangle \right)^3,
\end{equation}

\begin{equation}
\label{kurtdef}
 K=\frac{\mu_4}{\sigma^4} - 3, \quad \mu _4=\frac{1}{M} \sum_{i=1}^M \left( \frac{\delta n}{n}-\Biggl \langle \frac{\delta n}{n} \Biggr \rangle \right) ^4,
\end{equation}
where $\delta n = n- \langle n \rangle$, $n$ is the density, $M$ is the number of samples, $\langle ... \rangle$ denotes averaging over all samples, and $\sigma$ is the standard deviation of $\delta n/n$. Note that the relative density fluctuations $\delta n/n$ are defined as a ratio of the absolute density fluctuations and the instantaneous density since the BES signal is proportional to density. Both skewness and kurtosis were calculated as averages over four poloidal locations, and their poloidal variations in most cases were small.

\begin{figure}
    \centering
    \begin{subfigure}[b]{0.52\textwidth}
    	\includegraphics[clip,width=\textwidth]{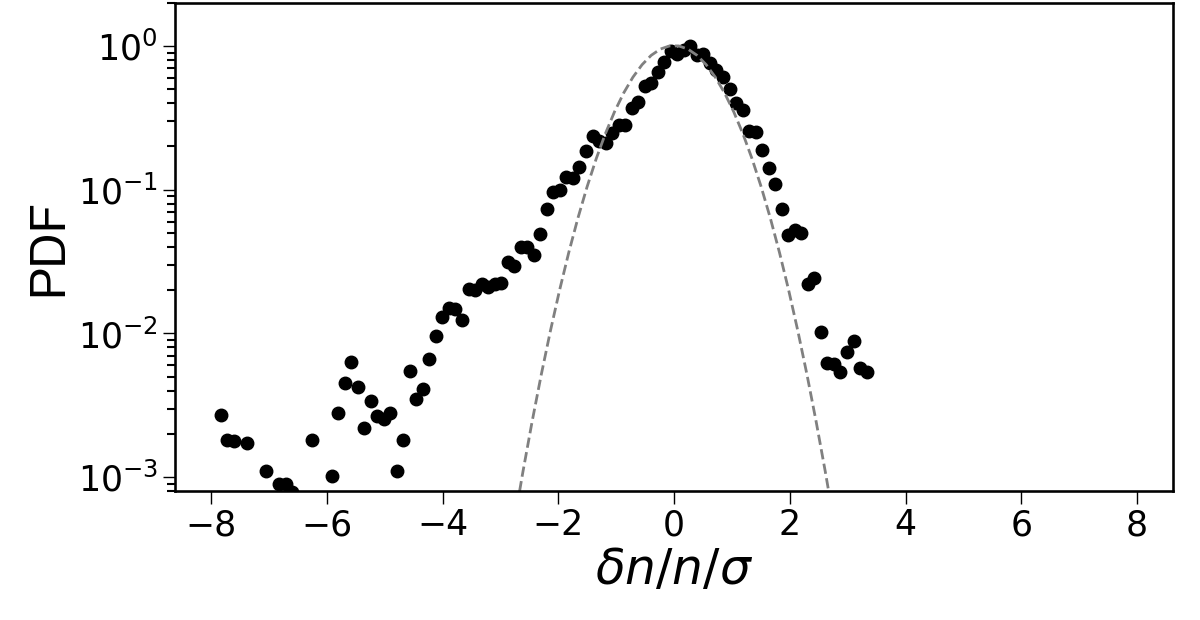}
	    \caption{}
    \end{subfigure}
    \hfill
    \begin{subfigure}[b]{0.52\textwidth}
    	\includegraphics[clip,width=\textwidth]{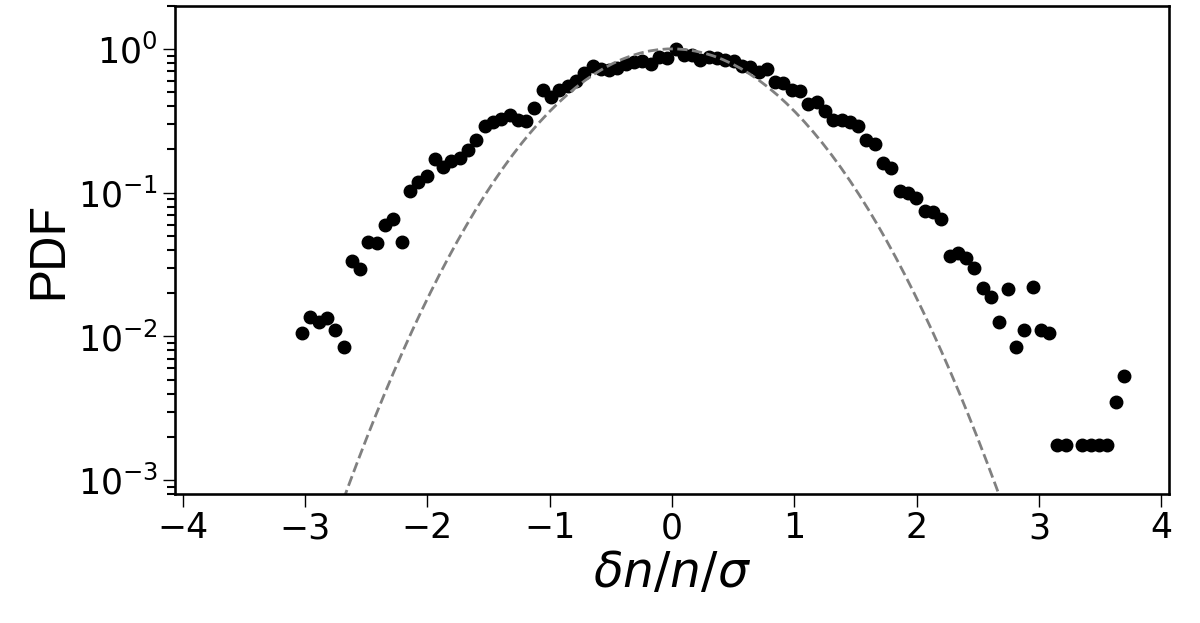}
    	\caption{}
    \end{subfigure}
    \hfill
	\begin{subfigure}[b]{0.52\textwidth}
    	\includegraphics[clip,width=\textwidth]{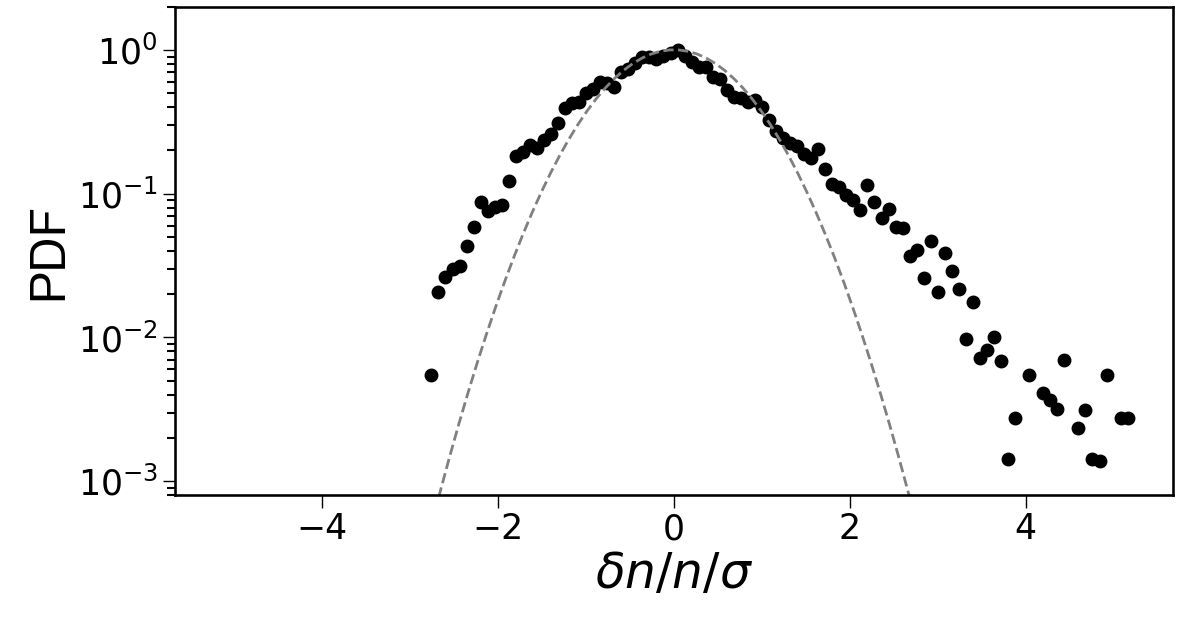}
    	\caption{}
    \end{subfigure}
    \caption{Probability distribution functions at three radial locations in the shot $\#$27292 for the data presented in Figure \ref{fig: timeseries}: (a) $R=1.21$ m, $\Delta R=-9 \quad \text{cm}$, $S=-1.13$, $K=3.92$; (b) $R=1.26$ m, $\Delta R=-4 \quad \text{cm}$, $S=-0.07$, $K=0.10$; (c) $R=1.28$ m, $\Delta R=-2 \quad \text{cm}$, $S=0.55$, $K=1.25$.}
	\label{fig: pdf}
\end{figure}

Figure \ref{fig: timeseries}(a) demonstrates that the relative density fluctuations at $\Delta R=-0.09$ m exhibit negative bursts with an amplitude above two standard deviations (indicated by the dotted line). The corresponding PDF of the density-fluctuation field is accordingly skewed towards negative values, as seen in Figure \ref{fig: pdf}(a). The skewness and kurtosis of the density-fluctuation field are $S=-1.13$ and $K=3.92$. The time trace of the density fluctuations at $\Delta R=-4$ cm demonstrates the presence of both positive and negative fluctuations in similar amounts, as can be seen from Figure \ref{fig: timeseries}(b). Figure \ref{fig: pdf}(a) shows that the corresponding PDF is symmetrical and has close to zero skewness and kurtosis. Closer to the separatrix, at $\Delta R=-2$ cm, the density-fluctuation field has large-amplitude positive bursts, as seen in Figure \ref{fig: timeseries}(c). The corresponding PDF is asymmetrical with a heavy positive tail and $S=0.55$ (see Figure \ref{fig: pdf}(c)). 

The deviation from Gaussianity of the PDFs of the density-fluctuation field is linked to the presence of large-amplitude, large-scale coherent structures $-$ blobs and holes $-$ at the edge of the outboard mid-plane plasma. Such structures are apparent from the 2-D images of the density-fluctuation field measured by the BES. These structures have high amplitudes $-$ up to $\delta n/n=40\%$ of the plasma density $-$ and span radially from 2 to 4 radial channels, i.e., 4 to 8 cm. 

\begin{figure}
    \centering
    \begin{subfigure}[b]{0.46\textwidth}
    	\includegraphics[trim={0 0 0 4cm},clip,width=\textwidth]{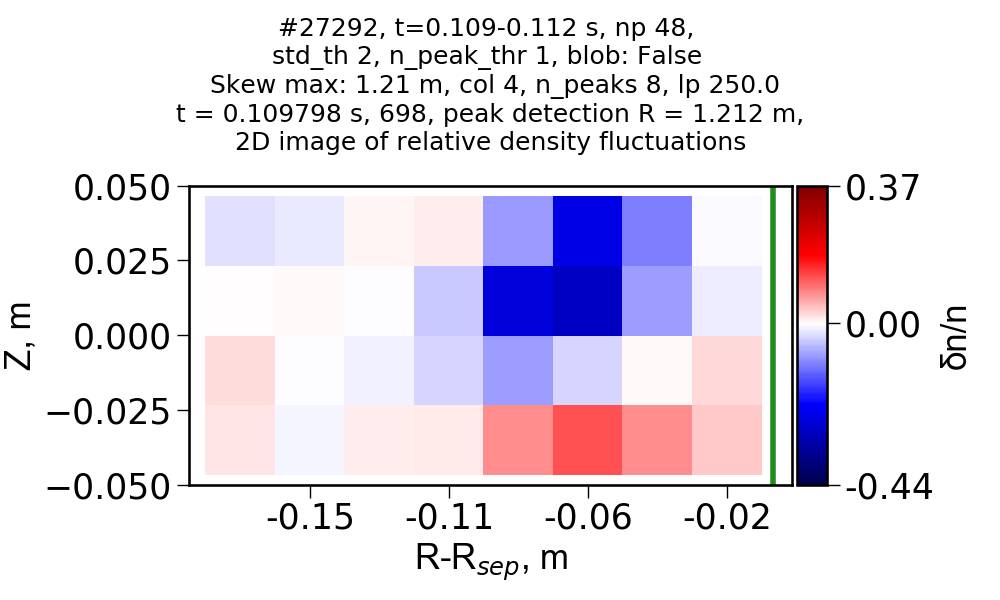}
	    \caption{$t_0=0.109798$s}
    \end{subfigure}
    \vfill
    \begin{subfigure}[b]{0.46\textwidth}
    	\includegraphics[trim={0 0 0 4cm},clip,width=\textwidth]{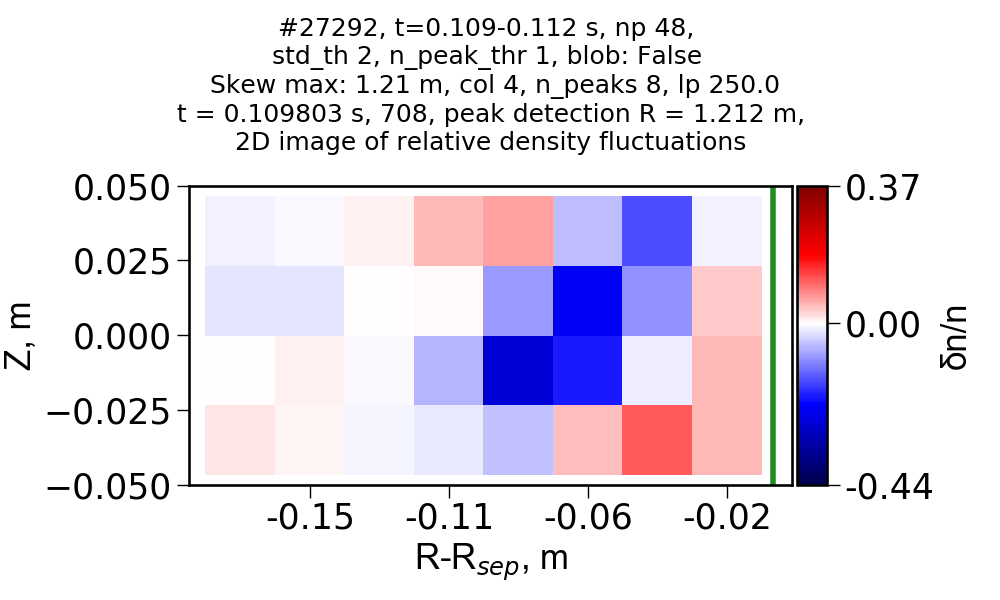}
    	\caption{$t_0+5 $ $\mu$s}
    \end{subfigure}
    \vfill
	\begin{subfigure}[b]{0.46\textwidth}
    	\includegraphics[trim={0 0 0 4cm},clip,width=\textwidth]{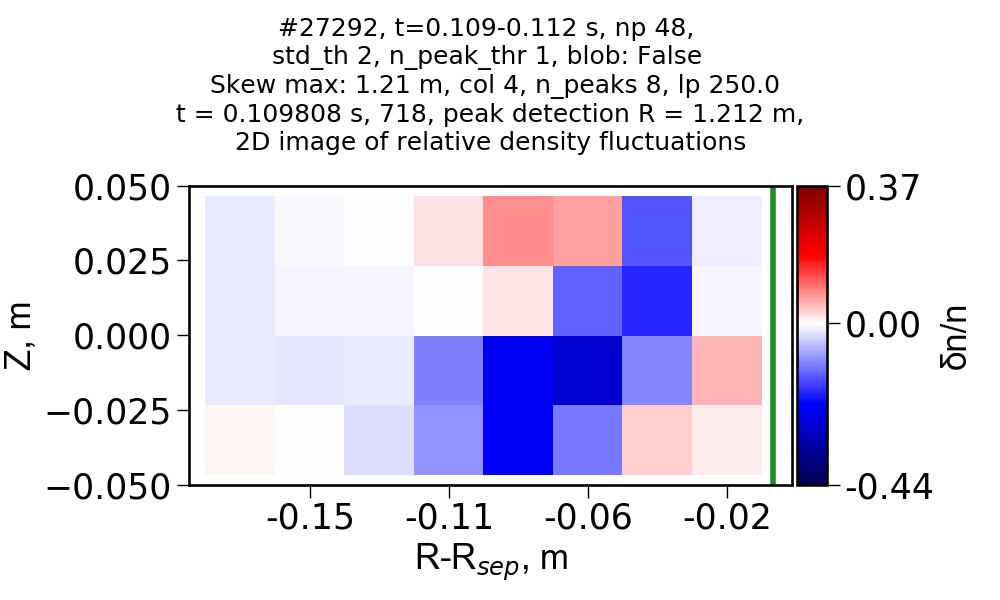}
    	\caption{$t_0+10 $ $\mu$s}
    \end{subfigure}
    \vfill
    \begin{subfigure}[b]{0.46\textwidth}
    	\includegraphics[trim={0 0 0 4cm},clip,width=\textwidth]{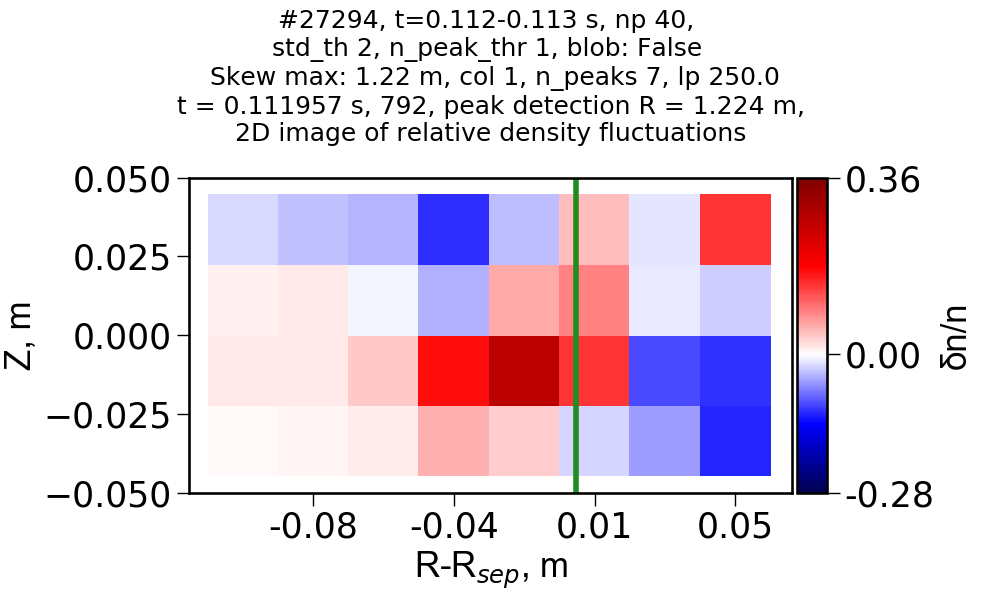}
	    \caption{$t_1=0.111967$s}
    \end{subfigure}
    \vfill
    \begin{subfigure}[b]{0.46\textwidth}
    	\includegraphics[trim={0 0 0 4cm},clip,width=\textwidth]{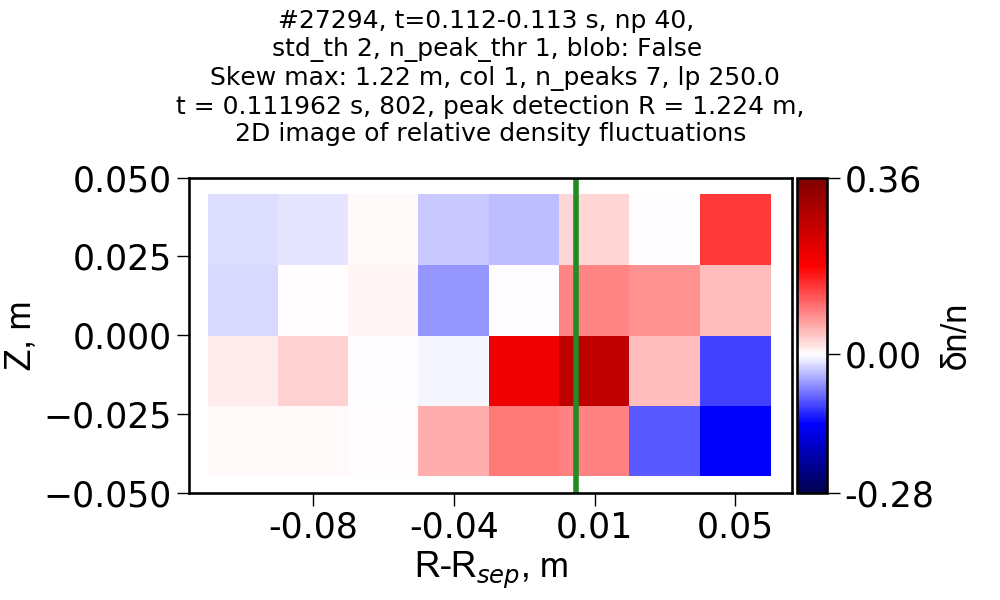}
    	\caption{$t_1+5 $ $\mu$s}
    \end{subfigure}
    \vfill
	\begin{subfigure}[b]{0.46\textwidth}
    	\includegraphics[trim={0 0 0 4cm},clip,width=\textwidth]{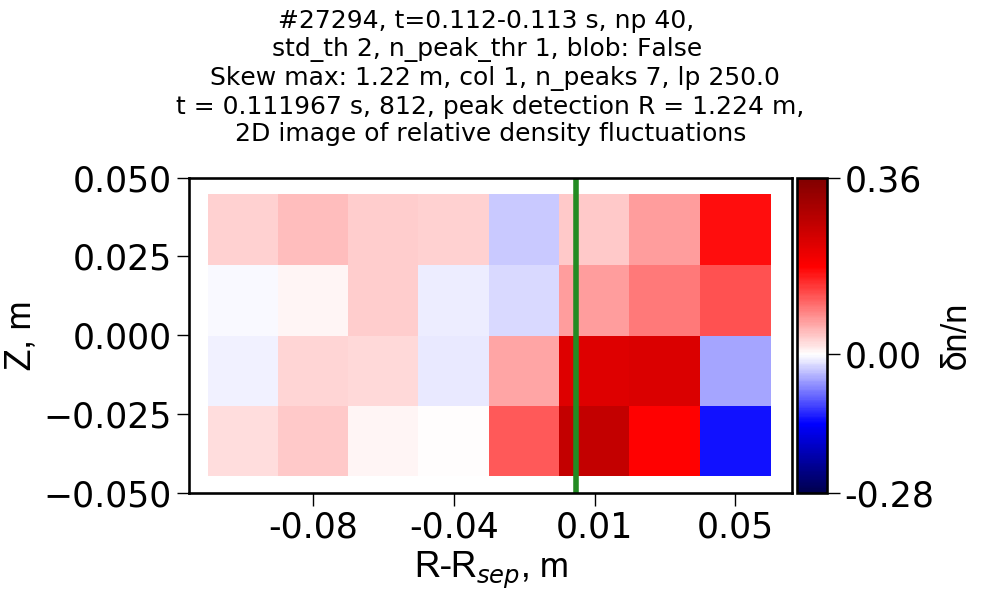}
    	\caption{$t_1+10 $ $\mu$s}
    \end{subfigure}
    \caption{Two-dimensional images of the density-fluctuation field: $(a)-(c)$ shot $\#$27292, $t=0.109798-0.109808$ s; $(d)-(f)$ shot $\#$27294 $t=0.111957-0.111967$ s; the time step between frames is 5 $\mu$s. The horizontal axis is the distance to the separatrix $\Delta R =R-R_{sep}$, the vertical axis $Z$ is the vertical coordinate. The green line is the location  of the separatrix.}
	\label{fig: snapshots}
\end{figure}

Examples of the 2-D density-fluctuation field during propagation of a density hole and a blob through the BES field of view are presented in Figure \ref{fig: snapshots}. The top and bottom rows represent two series of snapshots of $\delta n/n$. Each of the snapshot series is shown with a time step of 5 $\mu$s. In Figures \ref{fig: snapshots}(a)$-$(c), a density hole appears as passing downwards and radially inwards. Its poloidal velocity is equal to the apparent poloidal velocity originating from the equilibrium toroidal flow \citep{Ghim2012}. See section \ref{vpol_comp} for a comparison of poloidal velocity measured by BES and projection of toroidal velocity measured by charge exchange recombination spectroscopy (CXRS) on the poloidal plane. The 2-D images of the density-fluctuation field shown in Figures \ref{fig: snapshots}(d)$-$(f) exhibit a positive density burst, a blob, propagating radially outwards. It crosses the last closed flux surface (LCFS) while being advected poloidally. More discussion of the cross-correlation time delay estimation (CCTDE) algorithm used for the velocity calculations and radial dynamics of the density fluctuations is given in section \ref{Dynamics}.

The statistical properties of turbulence were analysed for a dataset containing shots and time ranges described in section \ref{Experim_setup}. The time traces of the relative density fluctuations were divided into time periods of the minimum duration of 0.5 ms. The turbulence characteristics were calculated for each of these time periods. The resulting dataset contained 745 points. 

\begin{figure}
    \centering
    \includegraphics[width=1\textwidth]{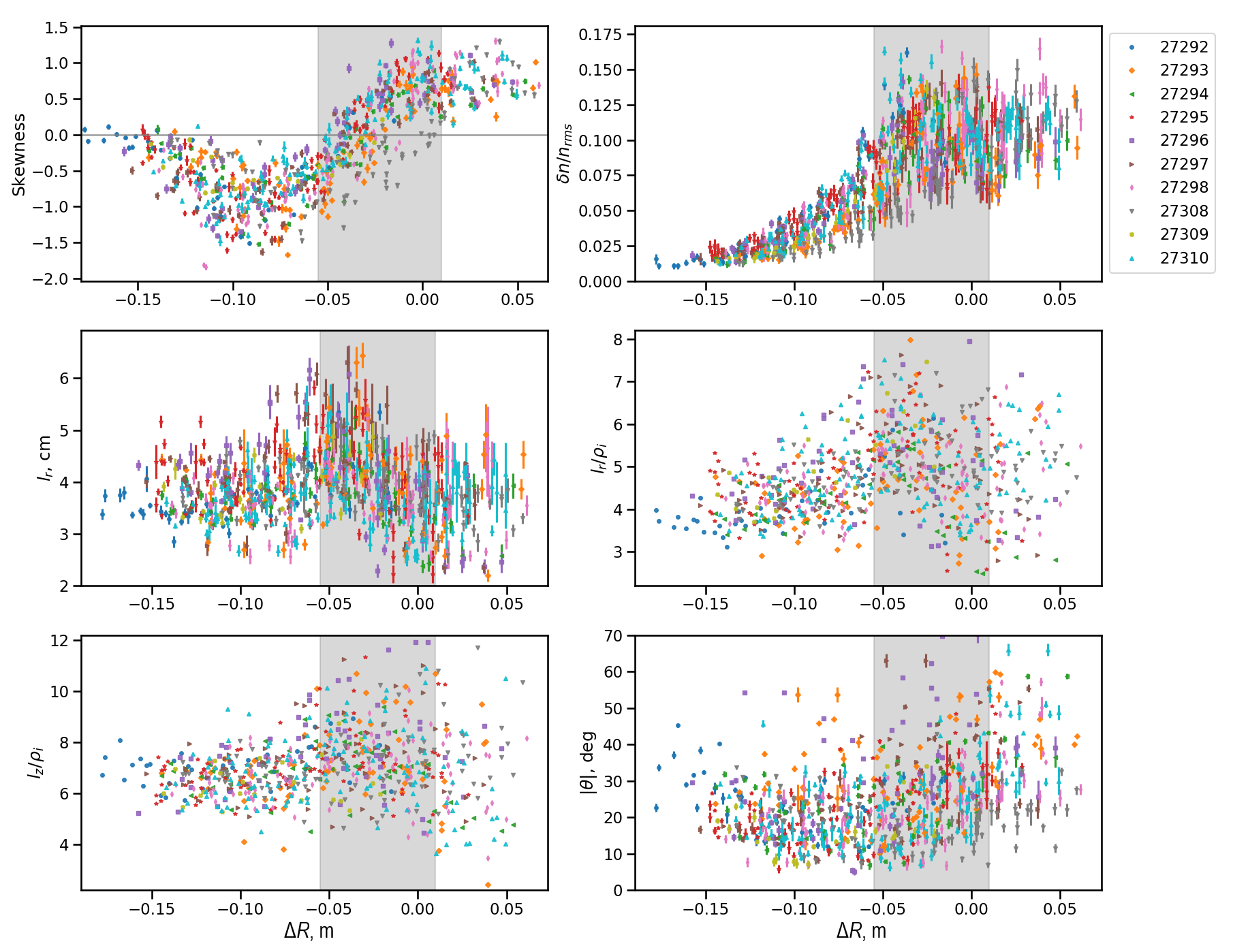}
    \caption{From left to right and top to bottom: skewness, root-mean-square (rms) value of the relative density fluctuation $\delta n/n_{rms}$, radial correlation length $l_r$ in cm, the same normalised to the ion Larmor radius $l_r/\rho_i$, the similarly normalised poloidal correlation length $l_z/\rho_i$, absolute value of the tilt angle of the 2-D spatial correlation function, all shown versus distance to the separatrix. The shadowed area is the radial range where the skewness takes values close to zero, i.e., the blob$-$hole formation region. The error bars in the skewness in this plot and later in the paper are the standard errors. The error bars in density fluctuations are the rms of $\delta n/n$ due to the photon noise \citep{Ghim2012}. The uncertainties in the correlation lengths and tilt angles correspond to the standard deviations of the fit to the 2-D correlation function (\ref{corr_rz}). The error bars in $l_{r,z}/\rho_i$ are omitted for simplicity due to the high errors in the ion-temperature measurements at the edge.}    
	\label{fig: sk_etc}
\end{figure}
Figure \ref{fig: sk_etc} shows the distribution of the turbulence characteristics versus the distance to the separatrix. One can see a striking similarity of the skewness profiles in all shots and time periods considered. There is a region inside the separatrix where the skewness changes sign from negative to positive in the direction of the LCFS. The asymmetry of the PDF becomes more pronounced as one moves away from the region of the zero skewness. A minimum in the skewness profile is at $7-10$ cm inside the separatrix. Beyond that point, at the locations towards the core, the symmetry of the density fluctuation field is gradually restored. Positive values of skewness up to 1.2 are observed in the edge and SOL. This suggests that the region where skewness changes sign from negative to positive corresponds to the formation zone where the blobs and holes are born. The region of zero skewness spans $0-5$ cm inside the separatrix, with the majority of points within $-$2 to $-$4 cm (Figure \ref{fig: sk_etc},  top left). The spreading of skewness and other variables reflects different plasma conditions in different shots and their change in time as the plasma parameters evolve in time. Moreover, the uncertainty in the position of the LCFS of approximately $ \pm 1$ cm can contribute to the scatter of the data. It should be noted that this picture does not exclude the possibility of the additional blob formation outside the LCFS, where instabilities can also drive the generation of the coherent structures \citep{Manz2015}. We will refer to the region of zero skewness as the 'blob$-$hole formation region', or 'birth zone', and to the region of minimum skewness as the 'hole-dominated region'.

The amplitude of the relative density fluctuations, estimated as their root-mean-square (rms) value, increases from $1\%$ at $\Delta R=-15$ cm to $17\%$ at $\Delta R=-2$ cm. The highest values are within the blob$-$hole formation region, with a slight decrease observed towards the separatrix where a strong equilibrium $E \times B$ flow shear is present (see section \ref{Equilibrium}). These data confirms the presence of strong density fluctuations at the edge of the L-mode MAST plasmas.

\subsubsection{Spatial properties of density fluctuations}
\label{spatial}
The correlation lengths of the density fluctuations were estimated by fitting a Gaussian function to the binned 2-D spatial cross-correlation function. The calculation of the binned correlation function is performed similarly to the method described by \citet{Fox2017}. Each cross-correlation function was first averaged over a time period and then fitted with the function in the form of
\begin{equation}
\label{corr_rz}
C(r, z)=p + (1-p)\exp \left[ -\left( \frac{r}{l_r} \right) ^2-\left( \frac{z}{l_z} \right)^2 \right] \cos(k_r r + k_z z),
\end{equation}
where $r$, $z$ are the radial and poloidal separations, respectively, between the reference and analysed channel, $l_r$ and $l_z$ are the radial and poloidal correlation lengths, $k_r$ and $k_z$ are the radial and poloidal wavenumbers, and $p$ is a constant that defines the offset in the cross-correlation function that may arise due to MHD modes  \citep{Ghim2012,Fox2017}. The tilt angle of the correlation function is defined as $\theta= \arctan (k_z/k_r)$.

The radial correlation length peaks towards the core of the plasma from the blob$-$hole formation region (Figure \ref{fig: sk_etc}, middle row). The high perpendicular background flow shear near the separatrix presumably plays an important role in shearing the large-scale positive fluctuations, thus reducing their radial size (see section \ref{Equilibrium}). Both $l_r$ and the normalised radial correlation length $l_r/\rho_i$, where $\rho_i$ is the ion Larmor radius, have maxima at $\Delta R=-(3-5)$ cm. The normalised radial correlation length varies from $l_r/\rho_i$=2.5 to 8. The poloidal correlation lengths also have an upshift near the blob$-$hole formation region with $l_z/\rho _i=4-12$ (Figure \ref{fig: sk_etc}, bottom left). Increased spatial scales of the density-fluctuation field support the presence of coherent structures near the region where the skewness changes its sign.

The tilt angles of the 2-D cross-correlation function of the density-fluctuation field were found to have minimal values near the hole-dominated region (Figure \ref{fig: sk_etc}, bottom right). An increase in the tilt angle towards the separatrix can be explained again by the presence of a strong equilibrium flow shear within that region. Smaller tilt angles within the hole-dominated region, as compared to the locations towards the core, support the presence of the larger-scale and longer-lived structures, which might survive shearing by either equilibrium or turbulence-driven oscillating flow shear. More details on the tilt angles of density fluctuations and shearing rates are presented in section \ref{Tilt_an}.

\subsubsection{Relationship between kurtosis and skewness}
\label{kurtosis_vs_skewness}
\begin{figure}
    \centering
   	\includegraphics[width=0.8\textwidth]{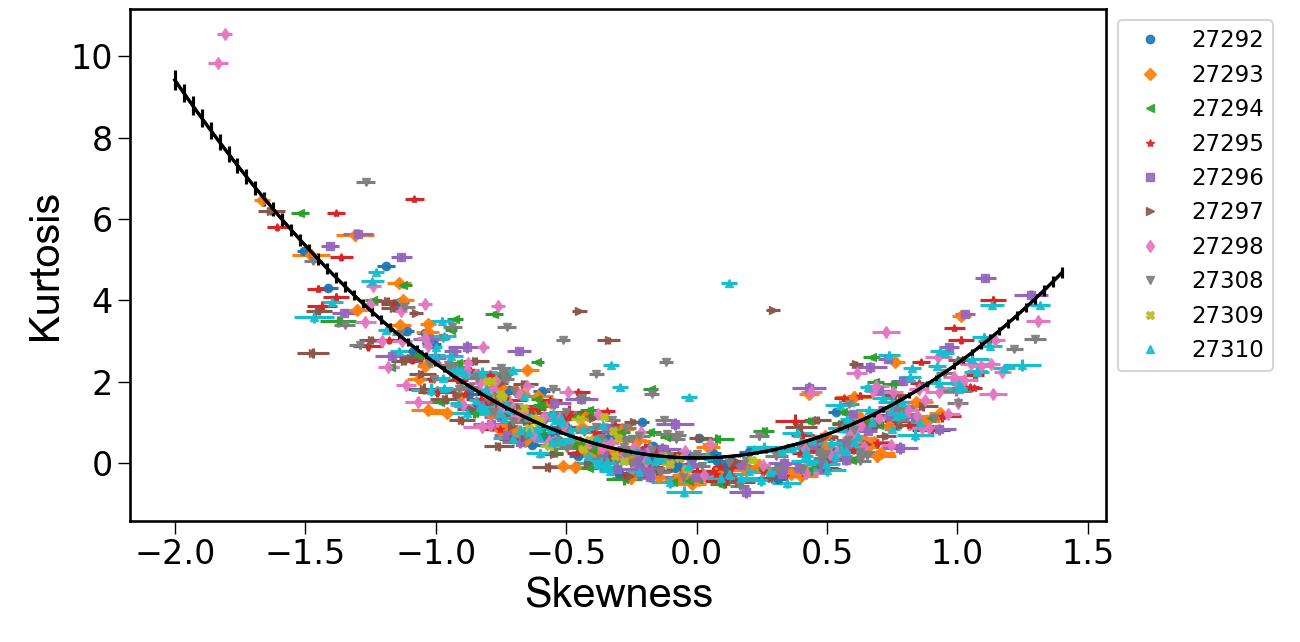}
    \caption{Kurtosis versus skewness. The black solid line is the second-order polynomial fit $K=(2.32 \pm 0.05) S^2 + (0.13 \pm 0.04)$.}
    \label{fig: kurt_sk}
\end{figure}

The relationship between the normalised third and fourth moments of the density fluctuations can provide useful insights into the underlying physical processes in the plasma \citep{Labit2007,Sattin2009,Garcia2012,Guszejnov2013}. Using the same dataset as Figure \ref{fig: sk_etc}, the kurtosis $K$ and the skewness $S$ are plotted against each other in Figure \ref{fig: kurt_sk}. The data was least-squares fitted using a second-order polynomial. There were no substantial difference in the quality of the fit with and without the linear term in the polynomial. The fit in the form of $K=(2.32 \pm 0.05) S^2 + (0.13 \pm 0.04)$ is shown by the black solid line. The error bars in the skewness and kurtosis are the standard errors.

Using mathematical models describing an intermittent signal, it has been shown that the presence of quadratic dependence of kurtosis on skewness is a manifestation of radially constant moments of intermittent events \citep{Garcia2012, Guszejnov2013}. The models relate the kurtosis and skewness of the fluctuating field as 
\begin{equation}
\label{ks}
K=\frac{M_2 M_4}{M_3^2} S^2 + C, 
\end{equation}
where $M_i$ is the $i$-th moment of the field describing intermittent events. In the model of \citet{Garcia2012}, $C$ is set to 1. In the model of \citet{Guszejnov2013}, $C=\langle N_{str} (N_{str}-1) \rangle  \langle N_{str} \rangle ^2$, with $N_{str}$ being the number of the events that are observed simultaneously. In the model of \citet{Garcia2012}, we assumed that a burst's waveform shape is characterised by a sharp rise and decay, consistent with the experimental observations. The model of \citet{Garcia2012} was reported to reproduce closely the $K-S$ relation obtained in experiments carried out in TCV \citep{Graves2005} and TORPEX \citep{Labit2007}; the model developed by \citet{Guszejnov2013} was shown to replicate the results from TORPEX \citep{Labit2007}. We tested the formula (\ref{ks}) against our experimental data. We calculated $C$ according to \citet{Guszejnov2013}. The moments $M_i$ were calculated using the extracted amplitudes of the density holes. Details on sampling the coherent structures are given in section \ref{Coherentstructures}. The number of the structures $N_{str}$ was estimated as the product of the linear density of density holes (see section \ref{n_hol_lin}) and the separatrix circumference. According to (\ref{ks}), the skewness$-$kurtosis relation should be $K=2.74S^2-0.07$, which is not far from the fit obtained based on our experimental data.

Agreement with the model (\ref{ks}) suggests that rare, weakly interacting (or non-interacting) coherent structures are transported radially leading to radially constant moments of the field of coherent structures. This supports the hypothesis that the density holes and blobs originate, as we argued above, up to several centimeters inside the separatrix, and spread radially towards the core and edge of the plasma, maintaining information about their distribution function. This transport of coherent structures results in negatively and positively skewed distribution functions of $\delta n/n$.

Note that the polynomial coefficients found in the fit for our dataset are similar to those obtained for H-mode NSTX plasmas, where the kurtosis dependence on skewness was $K=(2.28 \pm 0.10) S^2 + (0.19 \pm 0.2)$ \citep{Sattin2009}. The quadratic dependence of kurtosis on skewness of the edge density fluctuations has also been observed on ASDEX Upgrade \citep{Nold2010}.

\subsection{Coherent structures}
\label{Coherentstructures}

Non-Gaussian deviation of the PDF of the density-fluctuation field discussed in the previous paragraph is connected with the presence of relatively long-lived large-scale coherent structures. While numerous previous works \citep[see][and references therein]{DIppolito2011} have focused on studying blobs in the SOL of fusion devices, studies of density holes are quite rare. Here we will concentrate on the properties of density holes $-$ the coherent structures with less-than-average densities.

The study of these coherent structures was performed via the extraction of the negative density bursts with an absolute amplitude above the threshold of two standard deviations of the total density-fluctuation field. The characteristics of the density holes were analysed at the location of minimum skewness. The time periods during the propagation of each coherent structure through the BES field of view were between 10 and 60 $\mu$s, depending on the background flow velocity and the size of the structure. The amplitudes of the extracted relative density fluctuations corresponding to density holes were corrected for the photon noise.

\subsubsection{Correlation times and lengths of the structures}
\label{cor_time_len}

For each time series, the correlation times and lengths during the bursts of the negative density fluctuations and in between them were calculated. The calculation of the correlation times was done as follows. The density-fluctuation signals from poloidally separated channels were correlated in time, and the time delays between the maxima of the cross-correlation functions and the reference autocorrelation function were recorded. Since the predominant plasma motion appears to be in the poloidal direction, only poloidal channels were used to estimate the correlation times in the plasma frame. The obtained temporal cross-correlations were averaged over time and then fitted with the function 
\begin{equation}
\label{fit_ct}
C(t)=e^{-t/\tau_c}, 
\end{equation}
where $\tau_c$ is the correlation time.

\begin{figure}
    \centering
    \includegraphics[width=0.8\textwidth]{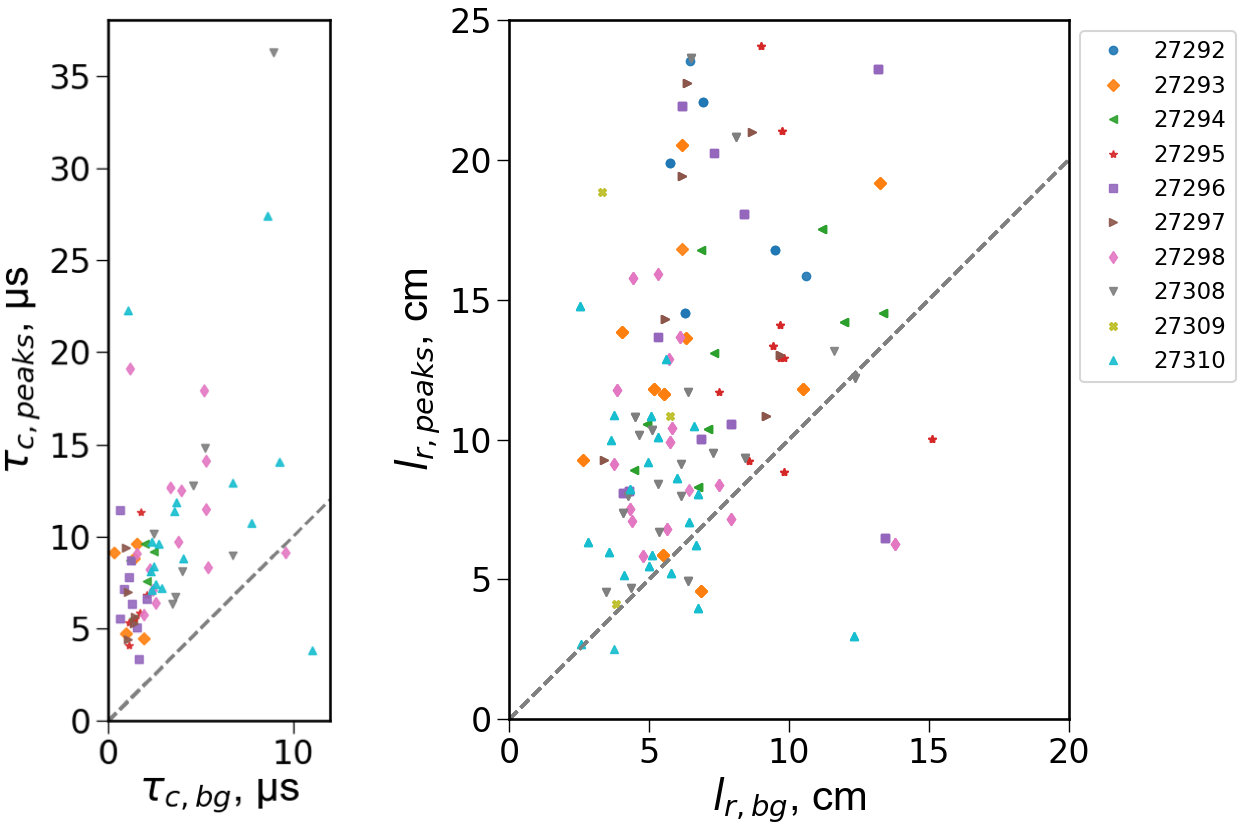}
    \caption{Correlation times (left) and radial correlation lengths (right) during bursts (vertical axes, subscript 'peaks') versus in between bursts (horizontal axes, subscript 'bg' for background). Each point corresponds to a separate time period.}
	\label{fig: ct}
\end{figure}

The correlation times of the high-amplitude under-densities were found to be higher than those of the density fluctuations in between the bursts, i.e., of the ambient turbulence. Figure \ref{fig: ct} (left) shows the correlation times of the density holes plotted against the correlation times of the background density fluctuations. Each point in the plot corresponds to correlation times for the holes and the background density fluctuations for a particular time period.

Figure \ref{fig: ct} (right) demonstrates that the density holes also have higher radial correlation lengths than the background fluctuations. A similar trend has been observed for the poloidal correlation length. The correlation lengths were estimated similarly to the way described in section \ref{spatial}. However, this time, the number of radial channels in the fit to (\ref{corr_rz}) was taken to be seven instead of three (as was done there), because the radial sizes of the coherent structures were larger. 

The uncertainties estimated from the standard deviations of the parameters of the fits were omitted from the plots to provide a clearer representation. The estimated uncertainties of the correlation times for the background turbulence are below $20 \%$. The errors in the correlation lengths of background turbulence are similar to those plotted for $l_r$ of the total density-fluctuation field in Figure \ref{fig: sk_etc}. The errors for the correlation times and lengths of the coherent structures are higher because of the smaller number of samples and are below $50 \%$.

Thus, the extracted high-amplitude events are larger-scale structures, and they are coherent in the sense that they have longer lifetimes than those of the ambient density fluctuations. 

\subsubsection{Number and amplitude of the structures} \label{n_hol_lin}
A characteristic feature of intermittent fluctuations is the number of observed structures per unit of poloidal length. As the number of the detected events in a time series depends on the advection velocity, the number of structures was normalised by the mean poloidal velocity. Thus, one can define a quantity that has units of inverse meters as the number of the observed events per time period per mean poloidal velocity of the structures: $N_{holes}=M /(\delta t v_{pol})$. This quantity can be regarded as the poloidal linear density of the structures. The density holes were observed to be advected by the same mean poloidal velocity as the total density fluctuation field. Details on the calculation of advection velocities are given in section \ref{cctde}.

\begin{figure}
    \centering
    \includegraphics[width=0.8\textwidth]{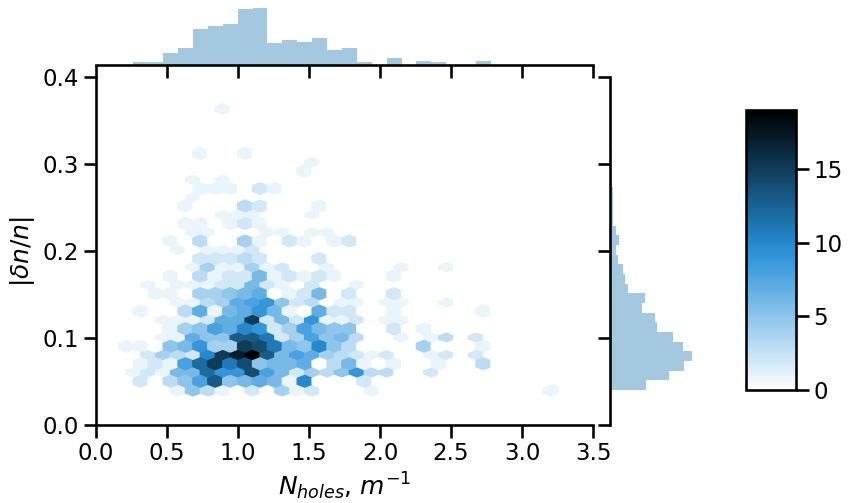}
	\caption{Distribution of the relative density fluctuation of holes versus the number of holes per poloidal length ($\delta t v_{pol}$). The colour shows the density of the points, the total number of points is 1025. The minimum time period over which the structures were counted was $\delta t =$1 ms. Poloidal velocity varied within 2-10 km/s, with the majority of the points having $v_{pol}=4-6$ km/s.}
    
	\label{fig: numpeak}

\end{figure}

The PDF of the linear density peaks at 1.1 $m^{-1}$, while the maximum of the PDF of $|\delta n/n|$ is at 0.1; see Figure \ref{fig: numpeak}. The average amplitude of the holes is approximately three times higher than the average $(\delta n/n)_{rms}=0.03$ of the total density-fluctuation field at the locations dominated by density holes (Figure \ref{fig: sk_etc}). The range of $N_{holes}$ corresponds to the existence of $1-5$ ($2-3$ in most cases) density-hole filaments at the low-field side of the tokamak plasma at a given moment in time. No correlation between the amplitude of the density holes and their number per unit length was observed. The events were rare enough so that no, or only weak, interaction between the coherent structures in the system could be present. 

The observed range of the linear densities of holes is similar to that of the blobs found in H-mode NSTX plasma: see, e.g., Figure 10 of \citet{Agostini2011}. Thus, there is a similarity both between the properties of the higher-order statistical moments (section \ref{kurtosis_vs_skewness}) and the number of structures per unit length in the spherical tokamaks MAST and NSTX.

\subsubsection{Tilt angles}
\label{Tilt_an}
\begin{figure}
    \centering
    \begin{subfigure}{0.46\textwidth}
        \centering
    	\includegraphics[width=\textwidth]{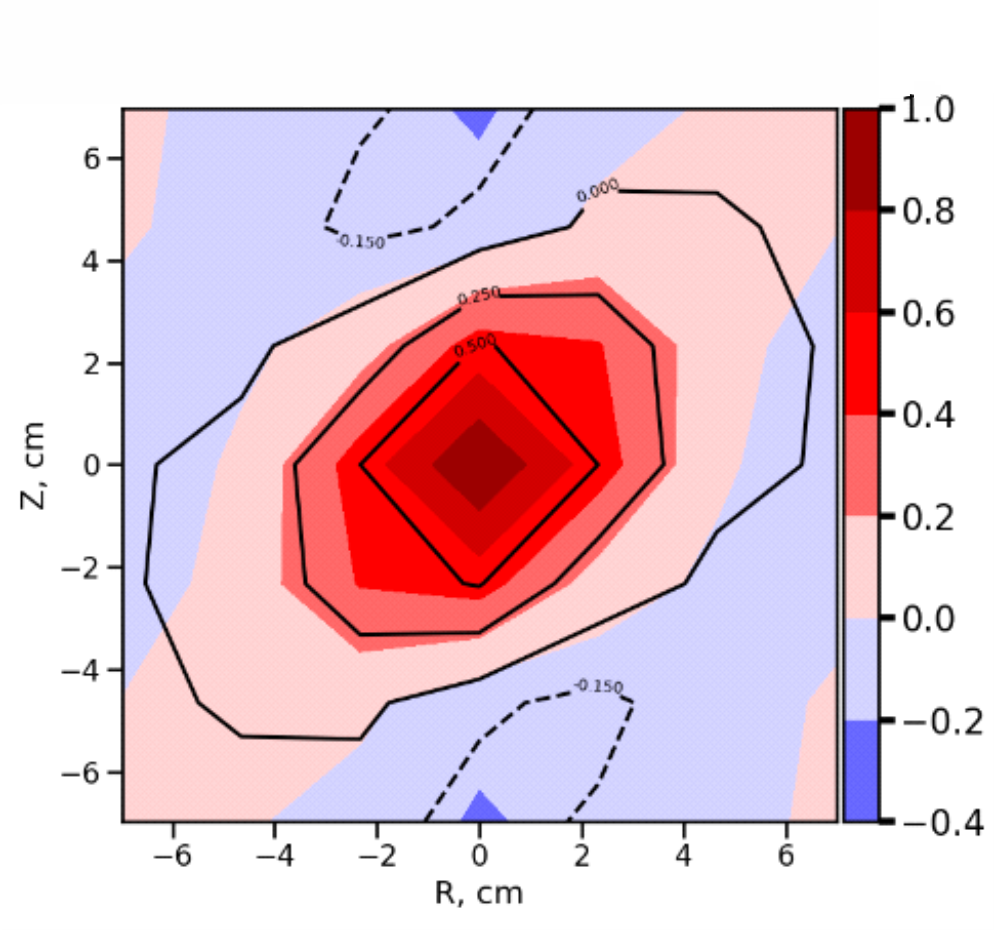}
        \caption{2-D cross-correlation function in between the bursts}
        \label{fig: 2-Dcc_between}
    \end{subfigure}
    \hspace{0.1cm}
    \begin{subfigure}{0.46\textwidth}
        \centering
        \includegraphics[width=\textwidth]{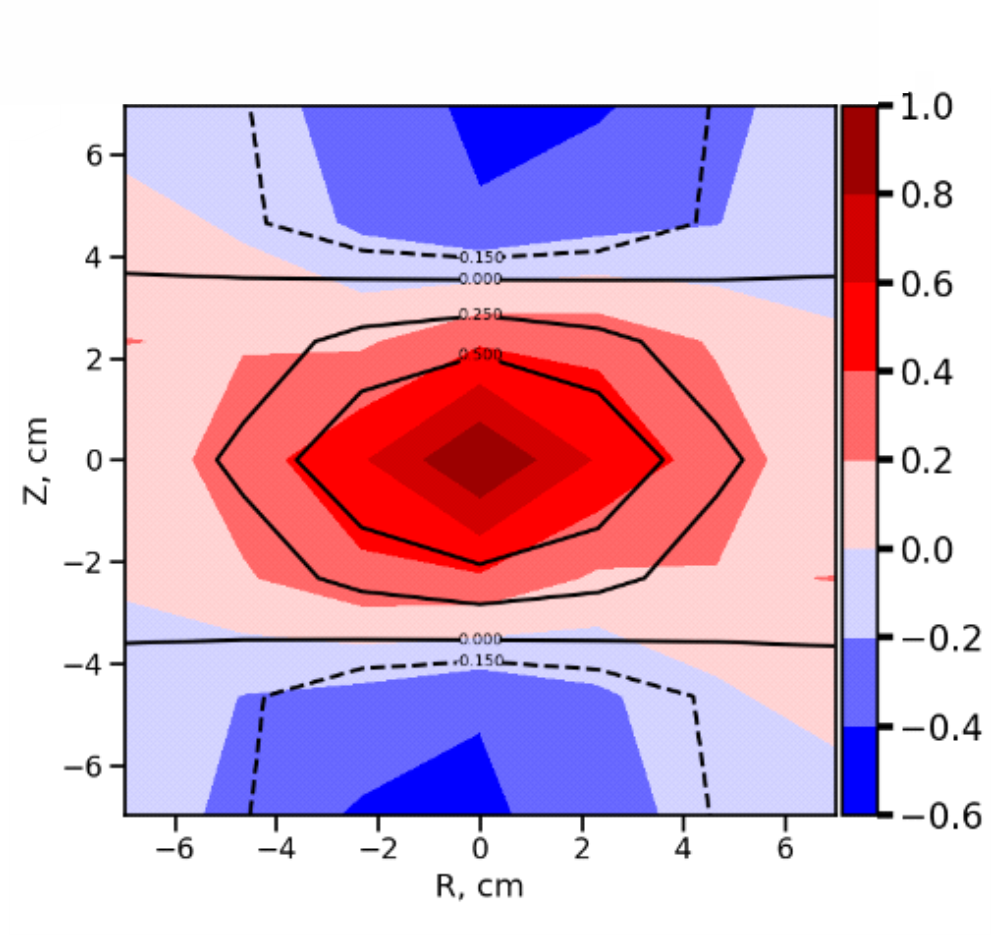}
        \caption{2-D cross-correlation function during the bursts}
        \label{fig: 2-Dcc_during}
    \end{subfigure}
    \caption{Conditionally averaged spatial two-dimensional cross-correlation function of density fluctuation field (a) in between and (b) during the high-amplitude negative bursts in $\delta n/n$. These data are for the shot $\# 27292$, $t=0.115-0.119$ ms. The cross-correlation function was calculated according to (\ref{corr_rz}).}
    
    \label{fig: tilt_angle}
\end{figure}

In many instances, we have observed that the tilt angles of the 2-D spatial cross-correlation functions were very different during the density-hole propagation and in between the high-amplitude events. Figure \ref{fig: tilt_angle} demonstrates this difference for the density fluctuations within the hole-dominated region at $R=1.16-1.23$ m, $\Delta R=-11$ to $-7$ cm. The density-fluctuation field had a small ($5 \degree$) tilt angle during the hole propagation and a tilt angle of $35 \degree$ in between the bursts. The tilting of the background turbulent fluctuations is associated with the presence of zonal flows, as will be argued in section \ref{Zonal flows}. The density-fluctuation bursts lead to the amplification of oscillating sheared flows that act on ambient turbulence \citep[see][]{Ivanov2020, Ivanov2022}.

\subsection{Relation of statistical properties to the equilibrium profiles}
\label{Equilibrium}

\begin{figure}
    \centering    
    \includegraphics[width=0.9\textwidth]{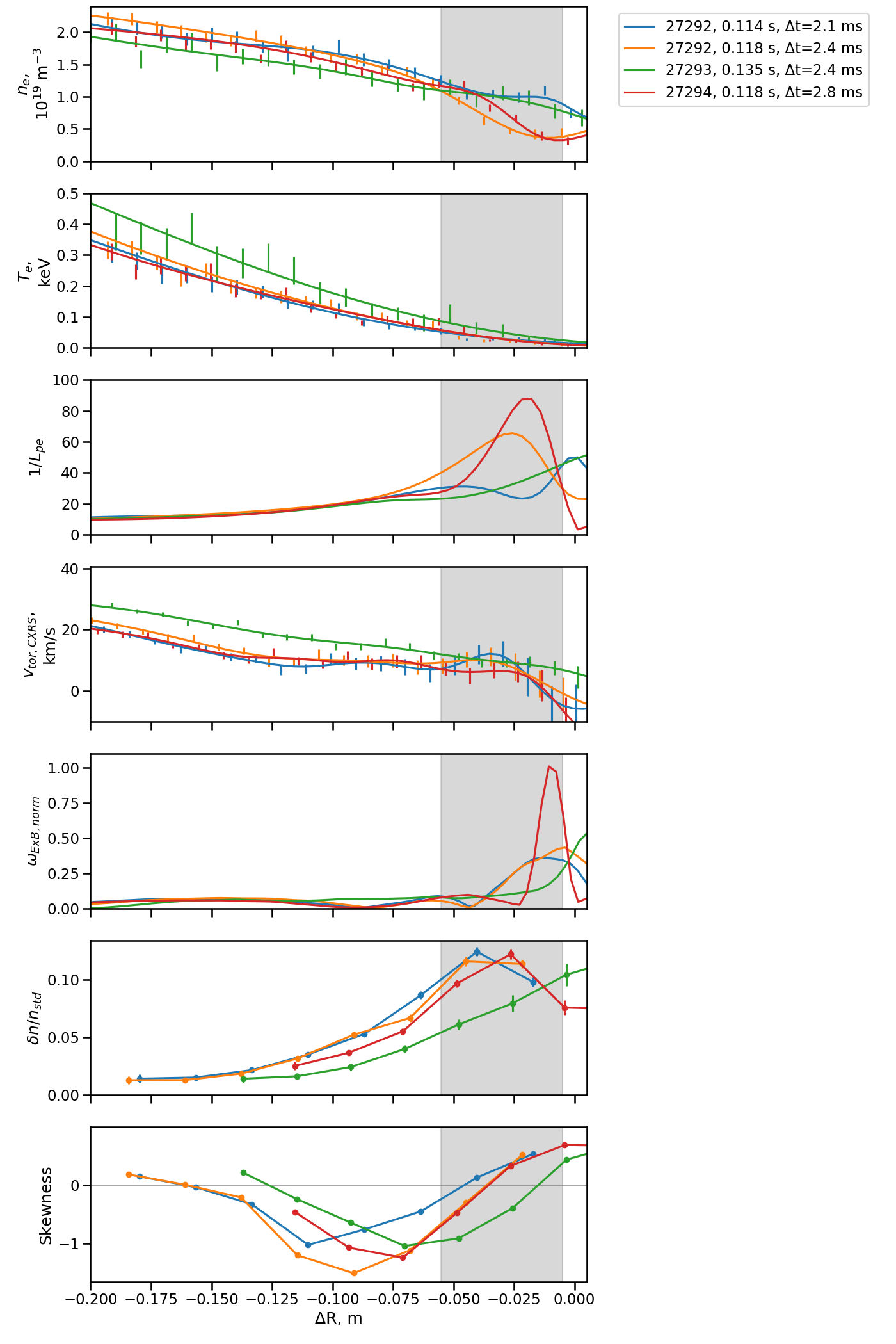}
    \caption{Top to bottom: radial profiles of electron density, electron temperature, normalised pressure gradient ($1/L_{pe}=|\nabla p_e/p_e|$), toroidal velocity, normalised perpendicular flow shear ($\omega _{E\times B, norm }=dv_{\perp}/dr (a/v_{th,i})$), standard deviation of $\delta n / n $, and skewness of $ \delta n / n $ against the distance to the separatrix $\Delta R=R-R_{sep}$ in shots $\#$27292 ($t=$0.114 s, 0.118 s), $\#$27293 ($t=$0.135 s), and $\#$27294 ($t=$0.118 s). The profiles are averaged over a time period $\Delta t$ indicated in the legend.}
    \label{fig: profiles_27292}
\end{figure}

Figure \ref{fig: profiles_27292} illustrates the time evolution of the profiles of various quantities plotted against the distance to the separatrix for three separate shots from the database. The skewness of the density fluctuations crosses zero and changes sign at different distances to the separatrix depending on the electron temperature and density profiles. The minimum of the skewness profile, where the density holes dominate the signal, also seems to follow the shift in the zero-skewness location. The relative-density-fluctuation level increases towards the edge and in most cases peaks near the blob$-$hole formation region.

We have found that, as plasma parameters evolve during each shot, the density-fluctuation levels and skewness profiles keep their shape with respect to the distance to the separatrix. The location where the skewness changes its sign is close to the position of the local maximum in the normalised electron pressure gradient $R/L_{pe}=R | \nabla p_e/p_e|$. This is in line with the simulations indicating formation of the blobs and holes as a result of the curvature-driven interchange instability \citep{Garcia2005,Russell2007}. Thus, the coherent structures are formed within the region of strong turbulence where the driving gradients of the kinetic profiles are maximal. The symmetry breaking of the density-fluctuation field is correlated with the region that is unstable to formation of blobs and density holes.

The background shear flow is responsible for shearing the turbulent eddies and coherent structures and reducing the amplitudes and radial sizes of the fluctuations. The normalised perpendicular background flow shear $\omega _{E \times B, norm}$ peaks a few centimeters towards the edge from the blob$-$hole formation region, at 1 cm inside the separatrix. The normalised flow shear shown in Figure \ref{fig: profiles_27292} is defined as $\omega _{E\times B, norm }=(\partial v_{\perp}/ \partial r) (a/v_{th,i})$, where $v_{\perp}=v_{tor} \sin \alpha$ is the perpendicular velocity, $v_{tor}$ is the toroidal velocity measured by CXRS, and $\alpha$ is the pitch angle of the magnetic field lines. As the shear increases towards the edge, the fluctuation amplitude  $\delta n / n $ drops. This is consistent with the observation of decreasing  $l_r/\rho_i$ towards the separatrix, shown in Figure \ref{fig: sk_etc}.

\subsection{Radial dynamics of edge fluctuations}
\label{Dynamics}
The general picture of the formation of coherent structures at the low-field side of magnetised plasma devices invokes charge polarisation inside blobs due to the curvature and $\nabla B$ drifts \citep{Krasheninnikov2001}. A density hole has an internal electric field in the opposite direction to that of a blob. The internal charge separation prompts the blob to move radially outwards and the hole radially inwards by means of the $E \times B$ drift. A similar mechanism has also been used for describing pellet deposition \citep{Parks2000,Rozhansky2004} and plasma jet penetration \citep{Rozhansky2006,Sladkomedova2018} in plasma fuelling experiments.

The dynamics of density holes can be described using the same theoretical concepts as usually applied to blobs \citep{Krasheninnikov2008}. In this work, we focused on radial dynamics within the confined region close to the separatrix at $r/a=0.85-0.88$. Hence it is expected that filamentary structures are disconnected from the sheath and plasma resistivity (hence perpendicular polarisation currents) plays an important role in the dynamics of coherent structures. Analytical theory \citep{Dippolito2002,Myra2006,Krasheninnikov2008,DIppolito2011} predicts radial velocities of blobs (or density holes) in the disconnected regime when inertial effects are important as follows:
\begin{equation}\label{eq_v_in}
 v_{r} = c_s \sqrt{\frac{l_r}{R} \bigg |\frac{\delta n}{n} \bigg |},
\end{equation}
where $c_s$ is the sound speed, $l_r$ is the characteristic size of a structure, and $R$ is the major radius. In this form, the equation captures the scaling of radial velocity with the square root of density perturbations as it is argued by \citet{DIppolito2011}. This formula is taken from \citet{Zweben2016}.

Radially propagating, poloidally localised, long-lived structures that exist in the presence of either equilibrium or zonal-flow shear have been observed in both gyrokinetic and fluid simulations of temperature-gradient-driven turbulence \citep{VanWyk2016, VanWyk2017, Ivanov2020, Ivanov2022}. The temperature dependence of the magnetic drifts employed in the fluid model implies that a local temperature perturbation can also give rise to plasma polarisation, which then, via the radial $E \times B$ drift, advects the temperature perturbation radially. Indeed, this is the basis of the self-advection mechanism of the aforementioned long-lived structures in the fluid model. As these are strongly nonlinear solutions that are kept alive by the linear drive, a balance of linear and nonlinear terms similar to (\ref{eq_v_in}) is also applicable to them. Therefore, the physics of propagating structures in the core and in the edge may not be dissimilar.

\subsubsection{Cross-correlation time-delay estimation method for velocimetry}
\label{cctde}
The radial and poloidal velocities of the density fluctuations were measured using cross-correlation time-delay estimation; our technique is similar to those described by \citet{Ghim2012,Cziegler2013,Fox2017}. Two signals are correlated in time and the time delay between the maxima of the cross-correlation function and the reference auto-correlation function is recorded. The velocities can be calculated using two or more points. For the calculation of poloidal velocities, up to four poloidal channels were used. We used an unbiased estimate of the cross-correlation function calculated using the fast Fourier transform and normalised to take values from $-1$ to $+1$. The algorithm was set to choose the best linear fit for the poloidal separations versus the time delays. Only statistically significant cross-correlations, estimated using the one-tailed t-test at a 0.05 significance level, were used in the fit. The minimum detectable poloidal velocity was $v_{p, min}=\Delta Z / \tau=2.3/16 \times 10^{4}=1.4$ km/s, where $\Delta Z$ is the poloidal separation between the two channels, and $\tau$ is the maximal time delay. In the case of the radial velocity, the cross-correlation was performed using two radial channels separated by 4.6 cm to minimise the effect of the correlation of the signals within the same density feature. The minimal radial velocity that the technique resolved was $v_{r, min}=2.8$ km/s. The effective time resolution of the calculated velocity time traces was 32 $\mu$s.

\subsubsection{Considerations of spurious radial velocities}
\label{Spur_vel}
\begin{figure}
    \centering
    \includegraphics[width=0.9\textwidth]{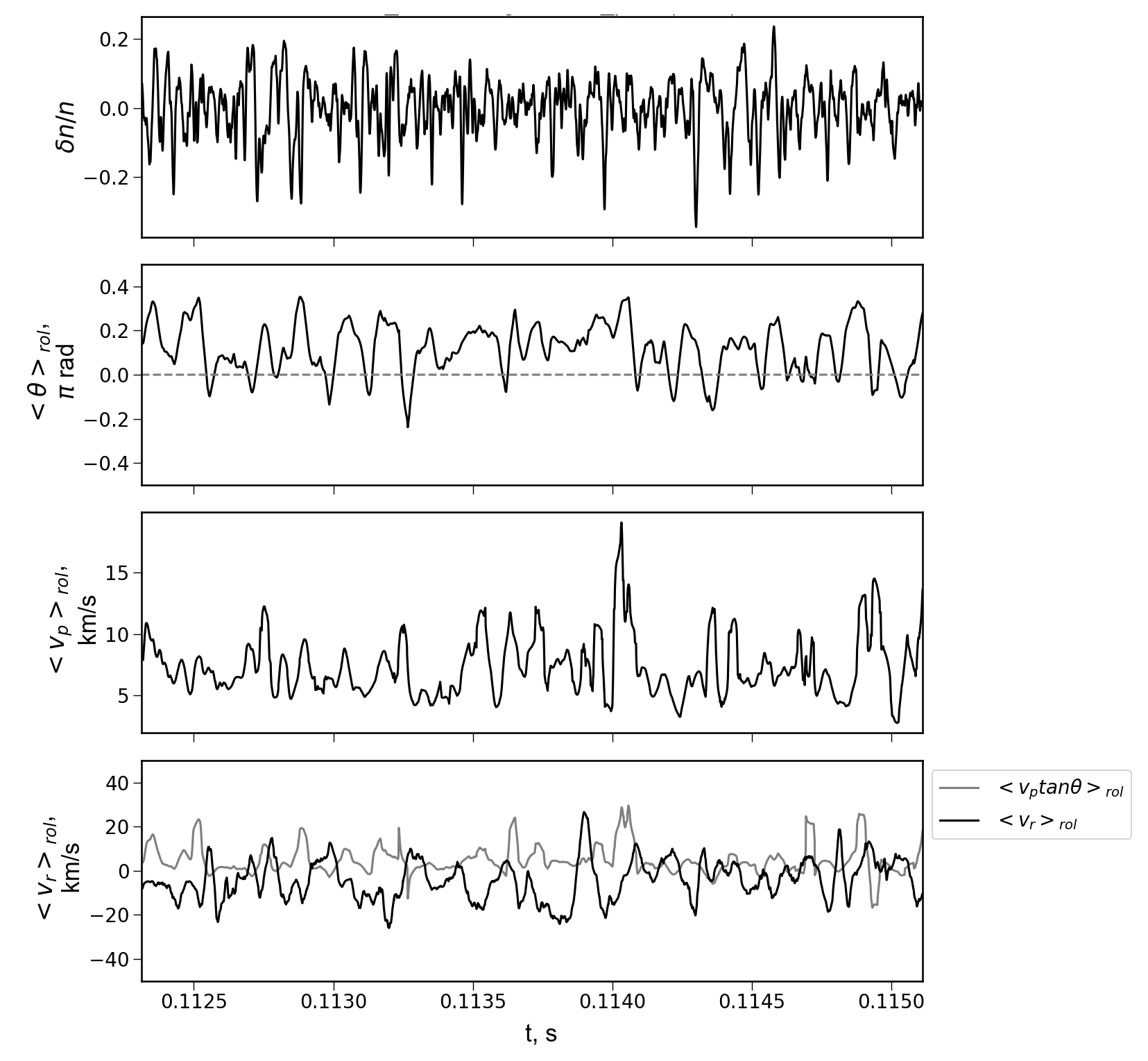}  	
    \caption{From top to bottom: time series of $ \delta n / n $, the tilt angle of the spatial correlation function, the poloidal velocity, the radial velocity for the shot $\#$27292, $R=$1.23 m, $\Delta R=-7$ cm. The grey line in the bottom plot is the product of the poloidal velocity measured by the BES and $\tan\theta$, where $\theta$ is the tilt angle. Here $<...>_{rol}$ denotes the rolling time averages over 64 points, i.e., over the period of 32 $\mu$s.}
    \label{fig: vrtilt_time}
\end{figure}

The density fluctuations are advected by the equilibrium toroidal flow predominantly driven by the momentum input from the NBI. Since the magnetic field lines are tilted, the toroidal motion (see $v_{tor, CXRS}$ in Figure \ref{fig: profiles_27292}) causes an apparent movement of the density fluctuations in the poloidal direction. Figure \ref{fig: vrtilt_time} shows the apparent poloidal velocity of the density fluctuations measured by the BES to be $4-15$ km/s in the ion diamagnetic direction.

If a structure is tilted, the apparent poloidal movement can result in spurious radial velocities even in the absence of any true radial motion. Let us consider here how it would affect the calculated velocities. First, the tilt angle, and poloidal and radial velocities were calculated with a time step equal to the time resolution of the BES of 0.5 $\mu$s. At each time step, using the cross-correlation method, these quantities were evaluated within the time window containing 64 temporal points. The radial velocities were calculated for each poloidal location and then the poloidal average was taken. A positive tilt angle and poloidal velocity (the ion diamagnetic direction is downwards) would result in a positive apparent radial velocity $v_p \tan \theta$, where $\theta$ is the tilt angle defined in the counterclockwise direction. The time series of the density fluctuations, the rolling averages of the tilt angle, and the poloidal and radial velocities for the shot $\#$27292 are shown in Figure \ref{fig: vrtilt_time}. The grey line in the bottom window indicates the apparent radial velocity originating from the poloidal movement of the tilted structures: $\langle v_p \tan \theta \rangle_{rol}$, where $\langle ... \rangle_{rol}$ is the rolling time average. The measured radial velocity and the effect from the tilt angle demonstrate different behaviour. Similar results were reproduced for other shots.

\begin{figure}
    \centering
    \includegraphics[width=0.7\textwidth]{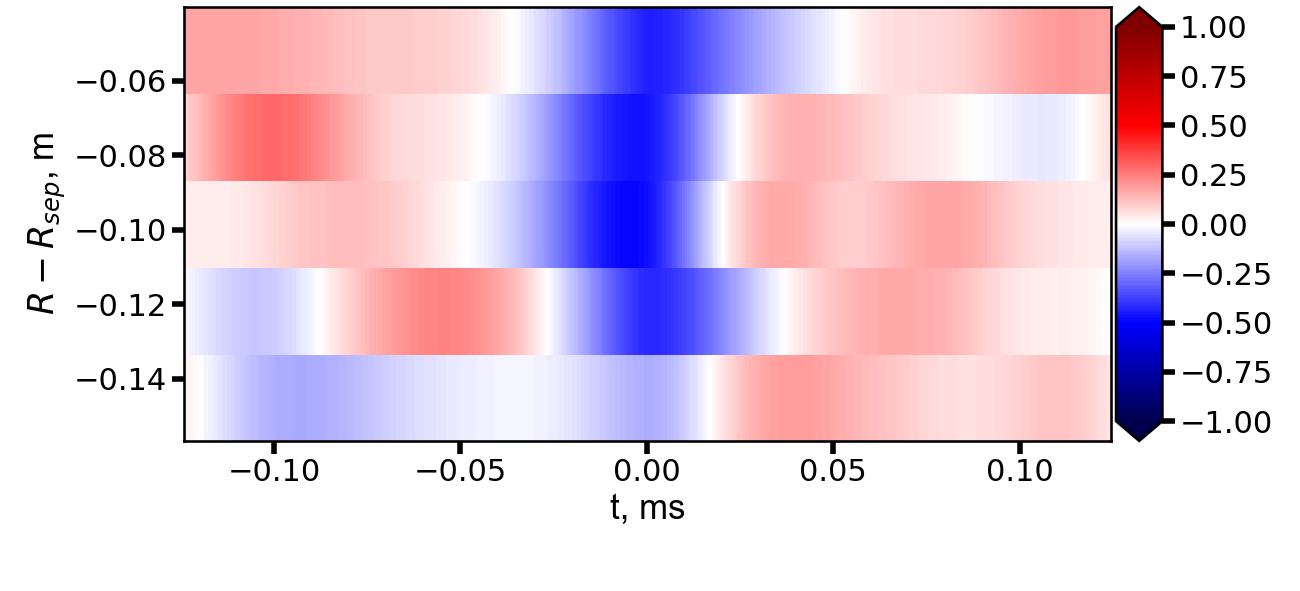}  	
    \caption{Temporal cross-correlation between the tilt angle and the radial velocity for different radial locations for the shot $\#$27292, $t=0.112-0.115$ s.}
    \label{fig: cor_th_vr_27292_t_112}
\end{figure}

A strong negative correlation coefficient between the tilt angle and the radial velocity was observed. If the effect of the apparent radial movement had dominated the estimate of the radial velocity, the correlation between the tilt angle and the radial velocity should have been positive. The temporal correlations between $\theta$ and $v_r$ for different radial locations during $t=0.112-0.115$ s are plotted in Figure \ref{fig: cor_th_vr_27292_t_112}. A negative correlation below $-0.2$ was observed for the radial locations on the outboard side of the plasma. The statistical significance of the correlation was confirmed by using the one-tailed t-test at a 0.05 significance level. The correlations were calculated using the same technique as described in section \ref{cctde}. A strong negative correlation between $\theta$ and $v_r$ was persistently observed during the time period of $0.110 - 0.120$ s across all radial locations. Thus, it can be concluded that radial velocities calculated using the method described above are not polluted by the apparent motion due to a combination of the tilting of the density features and toroidal rotation of the plasma.

The tilt angle can have a contribution from the instrument function, the so-called point-spread function (PSF): see \citet{Ghim2010} for more details. Increasing magnetic shear at the edge of the tokamak plasma results in the tilting of the field of view of each of the channels. Overlapping of the PSF between the neighbouring channels at the edge may cause spurious positive tilt angles of the correlation function \citep{Fox2017}. As the overlapping area between the PSFs of the diagonal and horizontal/vertical channels was similar, we have assumed this effect to be small and have not considered it in this work.

\subsubsection{Radial dynamics in the presence of coherent structures}
\label{rad_dyn_coh_str}

The negative correlation between the radial velocity and the tilt angle (Figure \ref{fig: cor_th_vr_27292_t_112}) indicates not only the fact that the calculated radial velocity is not a spurious effect caused by the poloidal movement of the tilted structures but also that the features that have stronger inward radial velocity have smaller tilt angles. This suggests that the density holes have larger inward velocities than the ambient turbulence.

\begin{figure}
    \centering
    \includegraphics[width=0.7\textwidth]{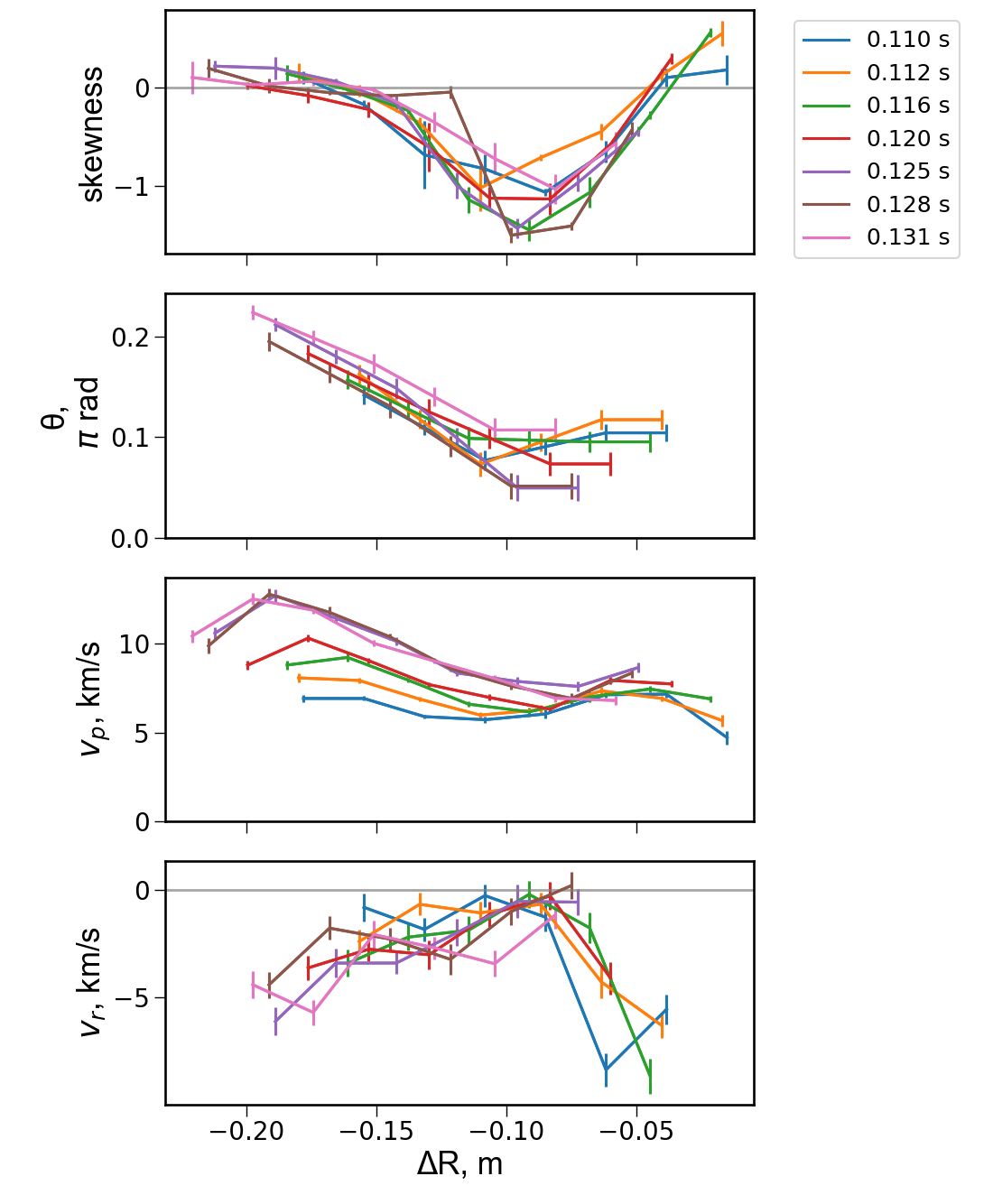}  	
    \caption{From top to bottom: radial profiles of the skewness of $ \delta n / n $, the tilt angle of the spatial correlation function, the poloidal and radial velocities for different time periods in the shot $\#$27292. The positive values of poloidal velocities are in the ion diamagnetic direction. The error bars for $\theta$, $v_p$, and $v_r$ are calculated as the standard errors of the mean for each quantity.}
    \label{fig: vr_27292}
\end{figure}

We have found that the radial velocities of the density fluctuations are highest in the region between the blob$-$hole formation region and the hole-dominated zone, with $v_r$ up to $-$8 km/s (see Figure \ref{fig: vr_27292}). The velocities calculated using the theoretical estimate (\ref{eq_v_in}) for the considered time moments in the vicinity of the blob$-$hole formation region at $\Delta R=-5$ cm, are $|v_{r}|=8-12$ km/s  and decrease towards the core, which is consistent with experimental observations. The experimental velocities do increase with the characteristic size and amplitudes of the density fluctuations, but a detailed comparison of the radial velocity measurements to (\ref{eq_v_in}) for various sizes and amplitudes of the structures has been left outside the scope of this paper. It should be noted that the radial velocities were measured for the total density-fluctuation field and included background turbulence fluctuations.

Propagation of density holes inward to the plasma core was observed on various other devices \citep{Boedo2003,Xu2009,Yan2011}. On NSTX, radial velocities of up to $-5$ km/s \citep{Boedo2014} were measured using Langmuir probes, close to the values obtained in this paper.  

The tilt angles are on average positive and have smaller values near the blob$-$hole formation and hole-dominated regions (Figure \ref{fig: vr_27292}). This is in line with the previous results covering the whole dataset (Figure \ref{fig: sk_etc}). As demonstrated in the previous section (Figure \ref{fig: tilt_angle}), the density holes have very small tilt angles. So the smaller tilt angle at the radial locations near the blob$-$hole birth zone and hole-dominated regions is due to the presence of coherent structures that survived shearing by either equilibrium or zonal flows.

\subsubsection{Comparison of poloidal velocities measured by BES and CXRS}
\label{vpol_comp}

\begin{figure}
    \centering
    \includegraphics[width=0.7\textwidth]{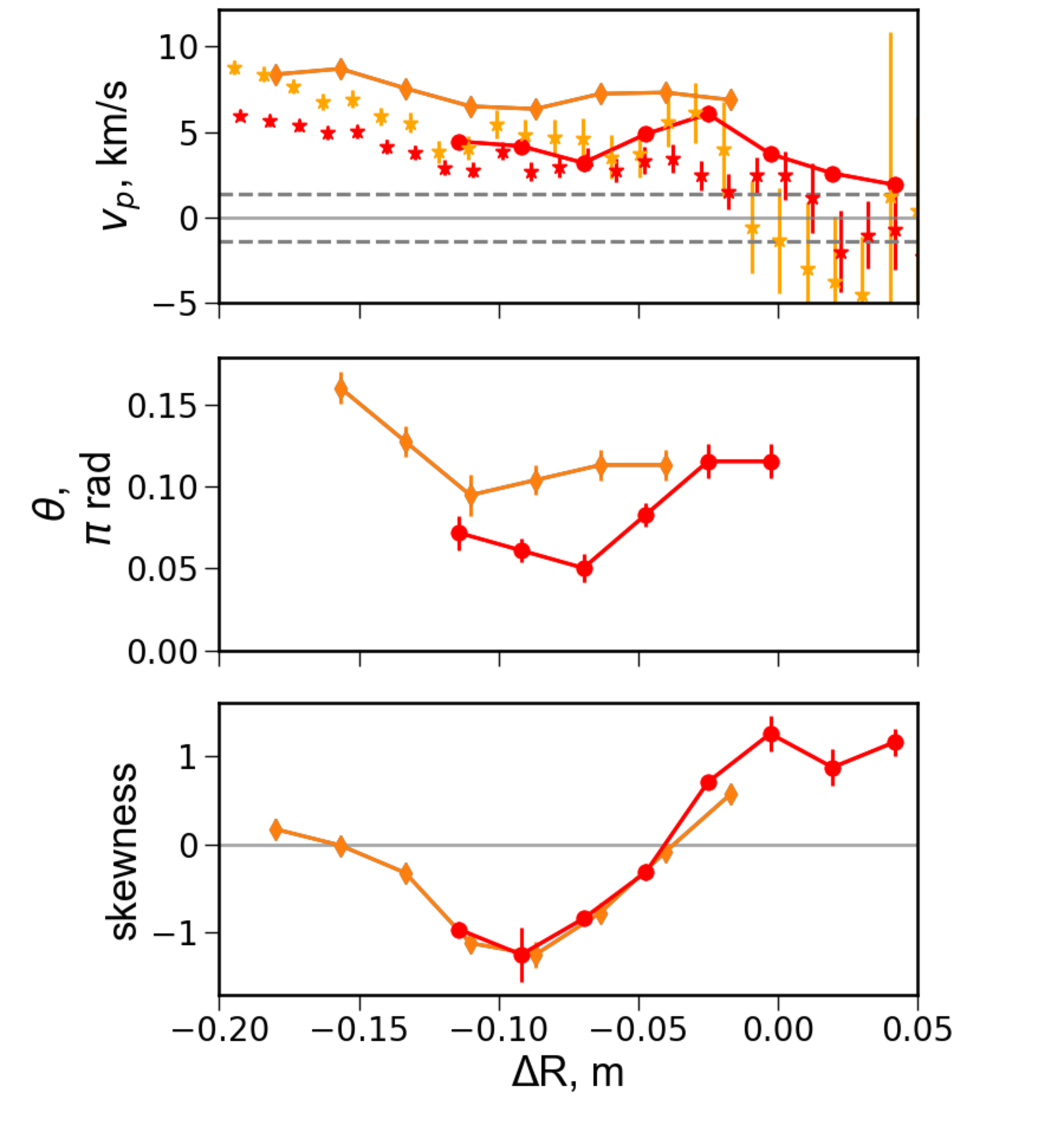}  	
    \caption{From top to bottom: radial profiles of the poloidal velocity, the tilt angle of the spatial correlation function, and the skewness of $ \delta n / n $. The orange colour and diamonds correspond to the shot $\#$27292, $t=0.1135-0.1185$ s, the red colour and circles to the shot $\#$27310, $t=0.1155-0.1205$ s. Stars indicate the velocities measured by the CXRS diagnostic. Grey dashed lines indicate the minimal poloidal velocity that can be measured.}
    \label{fig: vr_27292_27310}
\end{figure}

Let us look more closely at the performance of the CCTDE method for measuring the velocities by comparing it with the measurements of poloidal velocities (section \ref{cctde}) performed using the CXRS diagnostic. Figure \ref{fig: vr_27292_27310} shows a comparison of the time-averaged radial distributions of the poloidal velocities, tilt angles and skewnesses in two shots: $\#$27292 and $\#$27310. An interesting feature of both shots is that the skewness profiles plotted against the distance to the LCFS are very similar, with the region of the zero skewness close to the location of the peak normalised pressure gradient. Supporting the findings shown in Figure \ref{fig: sk_etc}, the tilt angles $\theta$ in the shot $\#$27310 increase towards the edge, where strong equilibrium flow shear is present.

Using the CXRS measurements, the apparent poloidal velocity was estimated as $v_{tor}\tan \alpha$, where $v_{tor}$ is the toroidal rotation velocity and $\alpha$ is the pitch angle of the magnetic field lines determined from the MSE measurements. The measurements done using the BES and CXRS are in fairly good agreement, with the former showing somewhat higher velocities. Any discrepancies between the poloidal velocities determined using the two diagnostics can be attributed to the uncertainties intrinsic in the measurements and the data-analysis techniques. Note that the poloidal velocity measured by the BES is a sum of the $E\times B$ velocity and phase velocity of turbulence: $v_{p}=v_{E\times B}+v_{ph}$. A more comprehensive estimate of the $E\times B$ velocity from the CXRS measurements would require measurements of the diamagnetic and neoclassical poloidal velocities. It should be noted that the BES measurements accordingly reflect different toroidal equilibrium velocities in two discharges.

\section{Zonal flows} 
\label{Zonal flows}

This section is devoted to the analysis of the poloidal flow dynamics of the density fluctuations. Poloidal velocities are calculated for all the shots in the available dataset; however, only two shots, $\#$27292 and $\#$27310, were characterised by sufficiently long time intervals for spectral analysis.

\subsection{Bursty dynamics of velocity fluctuations}
\label{v_vs_time}

\begin{figure}
    \centering
    \begin{subfigure}{0.71\textwidth}
        \includegraphics[width=\textwidth]{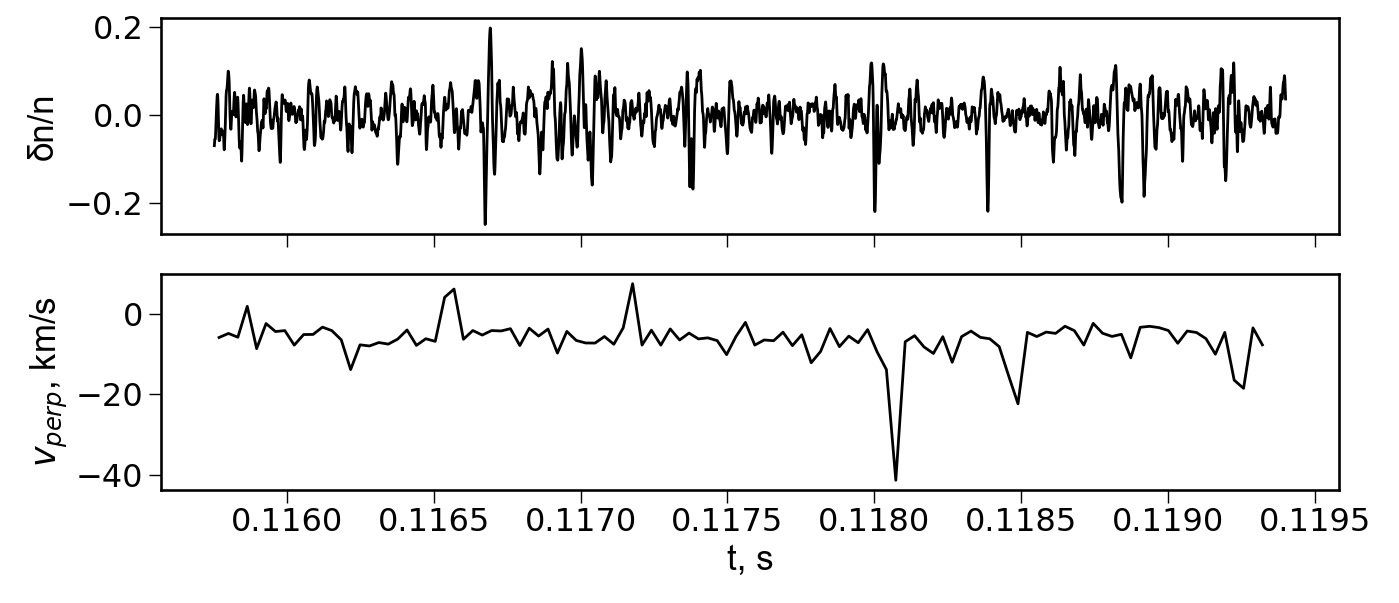}
        \caption{}
        \vspace{0.5cm}
    \end{subfigure}
    \vfill
    \begin{subfigure}{0.72\textwidth}
    	\includegraphics[width=\textwidth]{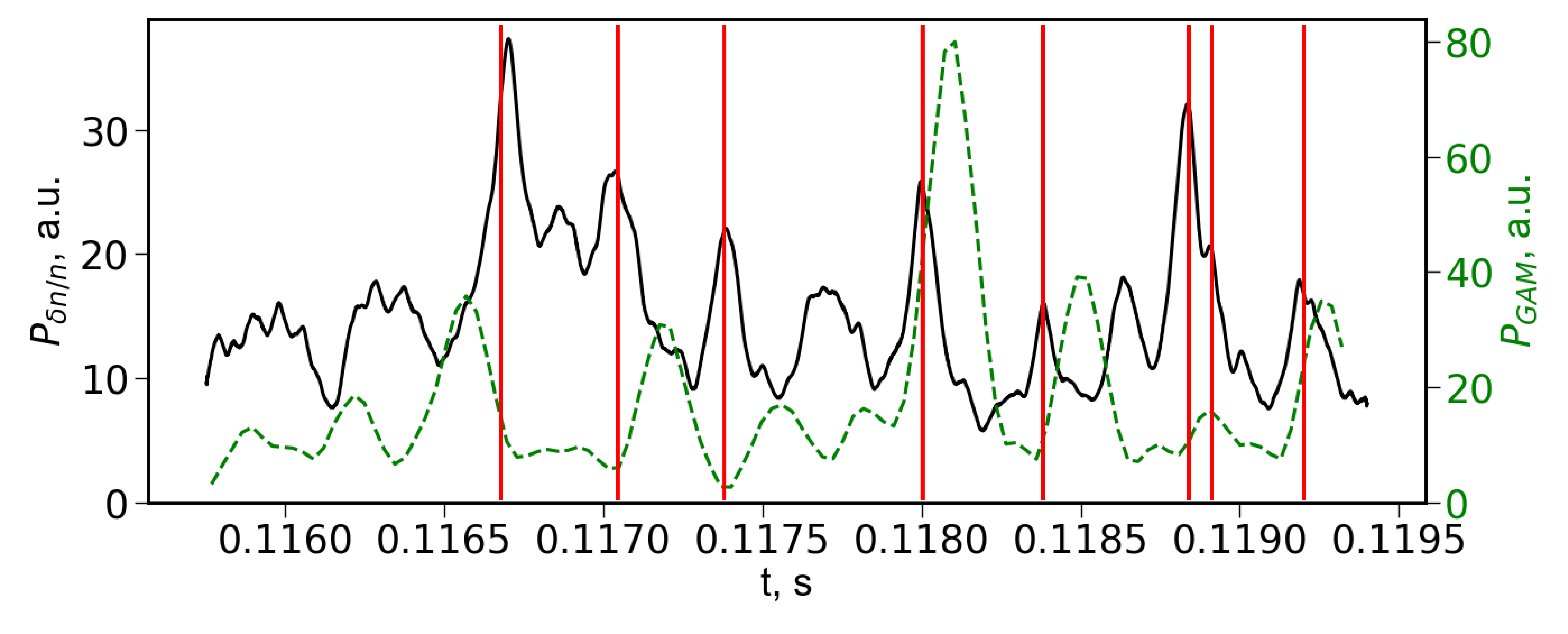}
        \caption{}
    \end{subfigure}
    
    \caption{Shot $\#27292$, $R=1.23$ m, $\Delta R=-7$ cm, $r/a=0.88$. (a) Time series of the relative density fluctuations $\delta n/n$ (top) and of the perpendicular velocity measured using the BES by the CCTDE method (bottom). (b) Wavelet power of $\delta n/n$ integrated over the frequency band 10$-$500 kHz (black solid line) and of the poloidal-velocity fluctuations at the GAM frequency integrated over 6$-$9 kHz (green dashed line). Vertical red lines denote the times when density holes are present at this location. A density hole is defined here as a negative density fluctuation with an amplitude above two standard deviations of the entire time series of $\delta n/n$.}
	
	\label{fig: wavelet_power}

\end{figure}

In all analysed shots, it is observed that the poloidal velocity fluctuations exhibit high-amplitude short bursts that usually follow a rise in the density-fluctuation power. An example of a time series of the density fluctuations and perpendicular velocity measured using the CCTDE method is presented in Figure \ref{fig: wavelet_power}(a). The perpendicular velocities have been estimated as $v_{\perp}=v_{pol} \cos \alpha$, where $v_{pol}$ is the poloidal velocity and $\alpha$ is the pitch angle of the magnetic field lines. As one can see from Figure \ref{fig: wavelet_power}(a), a reduction in $\delta n/n$ due to propagation of the density hole at $t=0.118$ s is followed by a sharp increase in the perpendicular velocity. A correlation between an increase in the amplitude of density fluctuations and perpendicular plasma flows suggests the generation of zonal flows as will be argued below.

The wavelet transform \citep{Daubechies1992} is a useful tool for analysing intermittent phenomena and is defined as the convolution of a wavelet and a given signal. We use the complex Morlet wavelet \citep{Morlet} for the calculation of the continuous wavelet transform of the density-fluctuation and velocity signals. The wavelets allow one to study fast dynamics of bursty signals. Figure \ref{fig: wavelet_power}(b) shows the time evolution of the wavelet power of $\delta n/n$, denoted $P_{\delta n/n}$, and of the perpendicular velocity fluctuations, $P_{GAM}$, for the shot $\#$27292. The wavelet power for density fluctuations was integrated over a frequency range of $10-500$ kHz, and for velocity fluctuations over $6-9$ kHz. The choice of the bandpass of the velocity fluctuations was dictated by the existence of coherent flow oscillations, the GAM, close to 8 kHz (see section \ref{mode}). Bursty dynamics of density and velocity fluctuations are manifested in the time series of $P_{\delta n/n}$ and $P_{GAM}$. High-amplitude short-scale bursts in $P_{\delta n/n}$ appear due to the propagation of density holes. The vertical red lines in Figure \ref{fig: wavelet_power}(b) correspond to the times when density holes are present in the signal. Recall that we define density holes as structures with negative $\delta n/n$ and an amplitude above two standard deviations. It is evident that a transient rise in the GAM power follows either an upshift in the power of broadband turbulent fluctuations or a burst due to a density hole. It should be noted that bursty behaviour of velocity oscillations has been previously found in MAST ohmic plasma using Langmuir probes, with the GAM frequency close to 10 kHz \citep{Hnat2018}.

Simulations of strong turbulence exhibit the generation of zonal flows by convective cells or blobs, with bursty behaviour of both density and perpendicular-velocity fluctuations \citep{Garcia2005,Garcia2003,Ivanov2020}. In particular, Figure \ref{fig: wavelet_power}(b) is reminiscent of the simulation results for ITG turbulence shown in Figure 13 of \citet{Ivanov2020}, with bursts in the heat flux followed by increases in the amplitude of sheared flows. 

\begin{figure}
    \centering
    \includegraphics[width=0.7\textwidth]{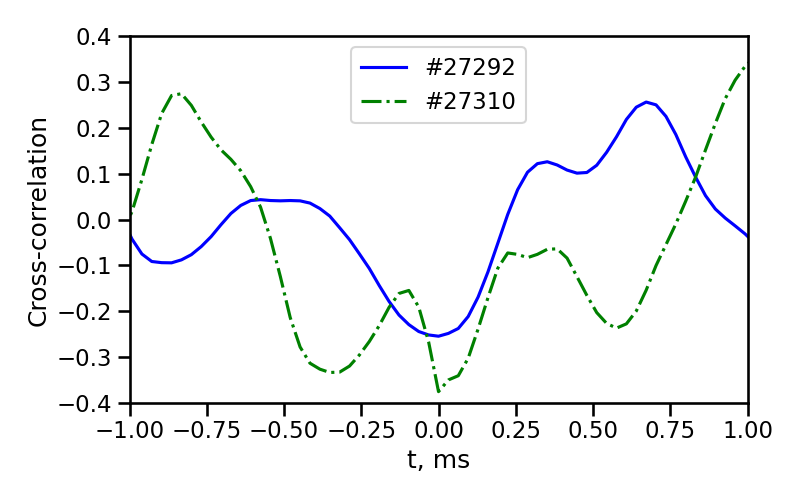}
    \caption{Biased temporal cross-correlation function between the wavelet power of GAM and the wavelet power of density fluctuations for the shots $\#27292$ $(t=0.115-0.1194$ s, $\Delta R=-7$ cm) and $\#27310$ $(t=0.1227-0.1252$ s, $\Delta R=-7$ cm). The statistically significant correlation coefficient at zero time delay, calculated using the one-tailed t-test at a 0.05 significance level, is 0.22. }
	
	\label{fig: ccwavelet_power}

\end{figure}

We have found that $P_{\delta n/n}$ and $P_{GAM}$ are anticorrelated (Figure \ref{fig: ccwavelet_power}). The cross-correlation coefficients at zero time delay were confirmed to be above the statistically significant correlation of 0.22, calculated using the one-tailed t-test at a 0.05 significance level. As the GAM power begins to rise before the peak in $P_{\delta n/n}$ is reached in certain cases, this alternating behaviour cannot be strictly labelled as anticorrelation. A similar picture was observed for other time periods in the same shots ($\#27292$ and $\#27310$), with the cross-correlation being negative at zero time delay and positive peaking at time delays below 0.1 ms. This supports the scenario of perpendicular-flow oscillations alternating with increases in density-fluctuation power associated both with broadband turbulence and with the propagation of large-amplitude large-scale coherent structures. Quasi-periodic bursty behaviour of density fluctuations and zonal flows was envisioned in predator$-$prey models that incorporated coupled dynamics of drift waves and zonal flows \citep{Malkov2001, Kim2003}. Our results suggest that coherent structures are part of a quasi-periodic self-regulating system that exhibits a behaviour similar to the standard framework of the drift wave-zonal flow turbulence: the kinetic energy is transferred from the fluctuations to sheared flows that suppress the initial drive and then decay allowing the perturbations to grow \citep{Diamond2005, Kobayashi2012, Ivanov2020, Ivanov2022}.

In section \ref{stat_fl}, we demonstrated that the generation and radial spreading of blobs and density holes are manifested in the skewed PDF of the density fluctuation field. Besides, we observed bursty behaviour in perpendicular velocity fluctuations, which is also reflected in the large higher-order moments of $v_\perp$. A comprehensive statistical investigation of the PDF of the velocity fluctuations would necessitate a larger dataset and is left for future studies.

\subsection{Mode identification}
\label{mode}

To identify zonal flows, we calculated the cross-coherence spectra of $\delta v_{\perp}=v_{\perp} - \langle v_{\perp}\rangle$ at two different poloidal locations near the mid-plane. The poloidal velocities were estimated using two channels for each poloidal location: two lower and two upper channels at $Z=(-1.15,-3.45)$ cm and $Z=(1.15,3.45$) cm. This gave us the velocities at the distances to the mid-plane of $Z= \pm 2.3$ cm. The velocities were evaluated at each time step by sliding the time window of the CCTDE technique. For the shot $\#$27292, the time period of $0.109-0.120$ s was used, which resulted in $> 7000$ time intervals, each of 1.5 ms duration. Most of the time periods partially overlapped with each other, and the total number of fully independent time intervals was four. The cross-coherence and cross-phase spectra were calculated using time-averaged cross- and auto-power spectra.

\begin{figure}
    \centering
     \includegraphics[width=0.7\textwidth]{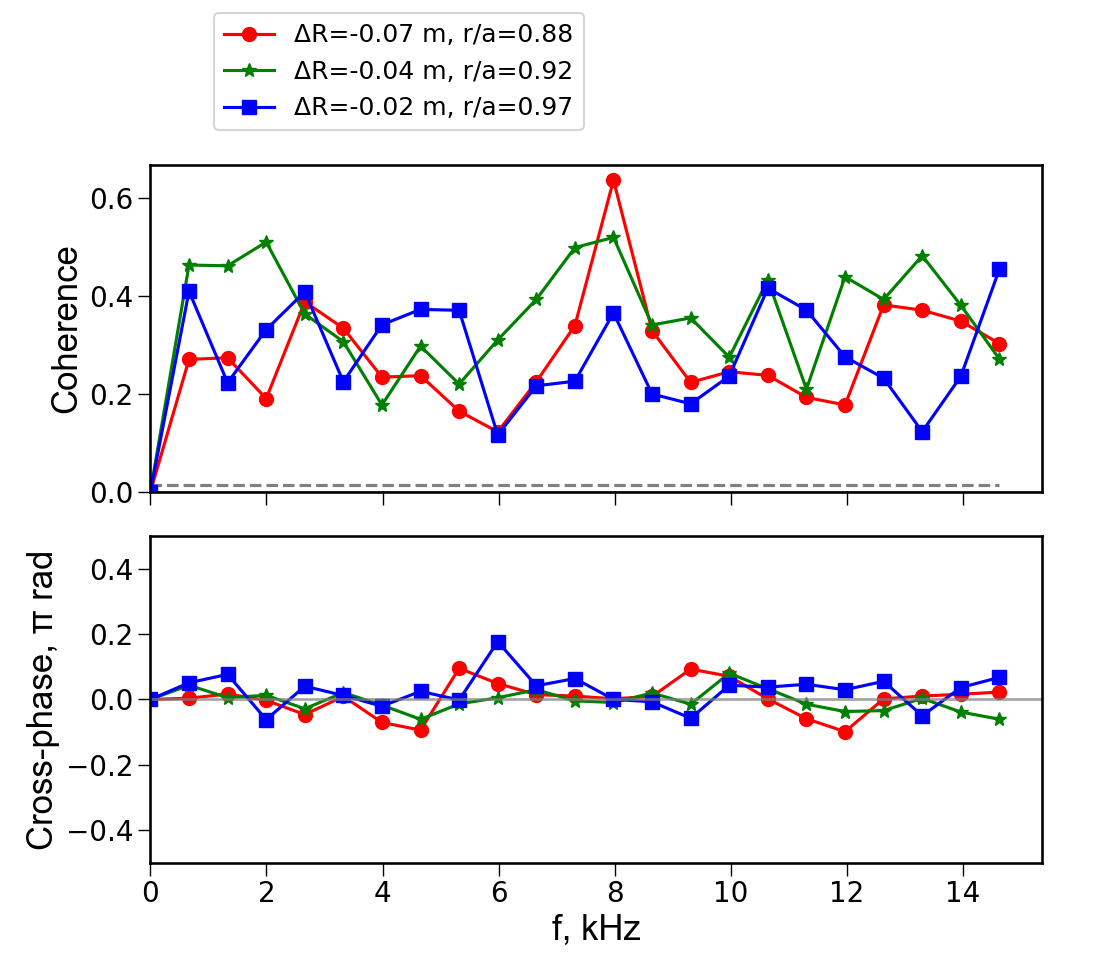}
    
	\caption{Cross-coherence and cross-phase spectra of $v_{\perp}$ between $Z=\pm 2$ cm at $R$=1.23 m ($\Delta R=-7$ cm), $R$=1.26 m ($\Delta R=-4$ cm) and  $R$=1.28 m ($\Delta R=-2$ cm) for the shot $\#27292$, $t=0.110-0.120$ ms. The grey dashed line indicates the noise floor.}
	\label{fig: cohvv}

\end{figure}

The cross-coherence spectra at three different radial locations are shown in Figure \ref{fig: cohvv}. A sharp peak in the coherence spectra at 8 kHz was observed at $\Delta R=-7$ cm ($r/a=0.88$) and a broader peak around the same frequency at $\Delta R=-4$ cm ($r/a=0.92$). A high degree of correlation between poloidally separated channels around a finite frequency is in agreement with the observed oscillations being poloidally symmetric coherent flows. The degree of cross-coherence decreases towards the edge and core of the plasma. However, the cross-phase between $v_{\perp}$ at two poloidal locations was close to zero not only for the presented locations but across all measurement locations.  

The GAM frequency for MAST conditions is estimated using the expression given by Eq. (20) of \citet{Gao2010}, which takes into account the shaping effects:
\begin{equation}\label{w_GAM}
\omega=\frac{v_{th,i}}{R} \sqrt{\left(\frac{7}{4}+\tau \right) \left( \frac{2}{k^2 + 1} \right)} \left[ 1 - s_k \frac{k^2}{4 k^2 + 4} - \epsilon ^2 \frac{9 k^2 + 3}{8 k^2 + 8} \right],
\end{equation}
where $v_{th,i}$ is the ion thermal velocity, $R$ is the major radius, $k$ is the elongation, $\epsilon$ is the inverse aspect ratio, $s_k=(r/k) \partial k / \partial r$, and $\tau=T_e/T_i$ \citep{Qiu2009}. We have neglected the Shafranov shift and other terms because they are negligible compared to those that have been retained here. This expression gives an estimate of $f= \omega / 2 \pi=18$ and $15$ kHz at $r/a=0.88$ and 0.92, respectively. Thus, the experimentally observed frequency of 8 kHz and the theoretical prediction of the GAM frequency for shaped plasmas are of the same order of magnitude.

\begin{figure}
    \centering
     \includegraphics[width=0.7\textwidth]{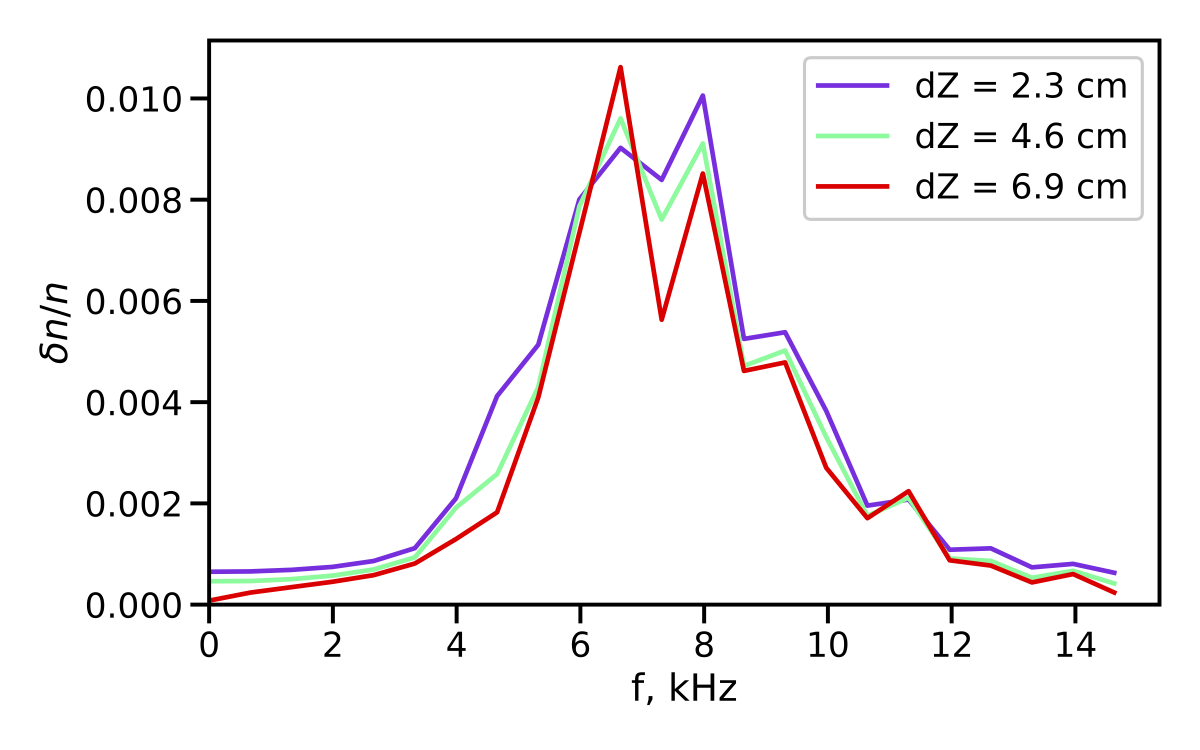}
    
	\caption{Cross-power of $\delta n/n$ at different poloidal locations for the shot $\#27292$, $t=0.110-0.120$ ms, at $\Delta R=-7$ cm.}
	\label{fig: crosspower_nn}

\end{figure}

The distinction of the higher-frequency sheared flows from the low-frequency (LF) zonal-flow branch is the presence of the oscillation at approximately 8 kHz in the $\delta n/n$ spectra. The peaking at the GAM frequency in the cross-power spectra of $\delta n/n$ between the poloidally separated channels was found in both shots analysed here at all viewing radii. The amplitude of the perturbation at the GAM frequency was $\delta n /n \sim 1 \%$ (see Figure \ref{fig: crosspower_nn}). This constitutes approximately $10\%$ of the total level of density fluctuations at $\Delta R=-7$ cm (see Figure \ref{fig: profiles_27292}). The cross-phase relation between the signals at the GAM frequency corresponds to the poloidal mode number $m=10$, which is higher than predicted by the standard theory $-$ the $m=1$ density perturbation \citep[see][]{Diamond2005}.

\begin{figure}
    \centering
     \includegraphics[width=0.7\textwidth]{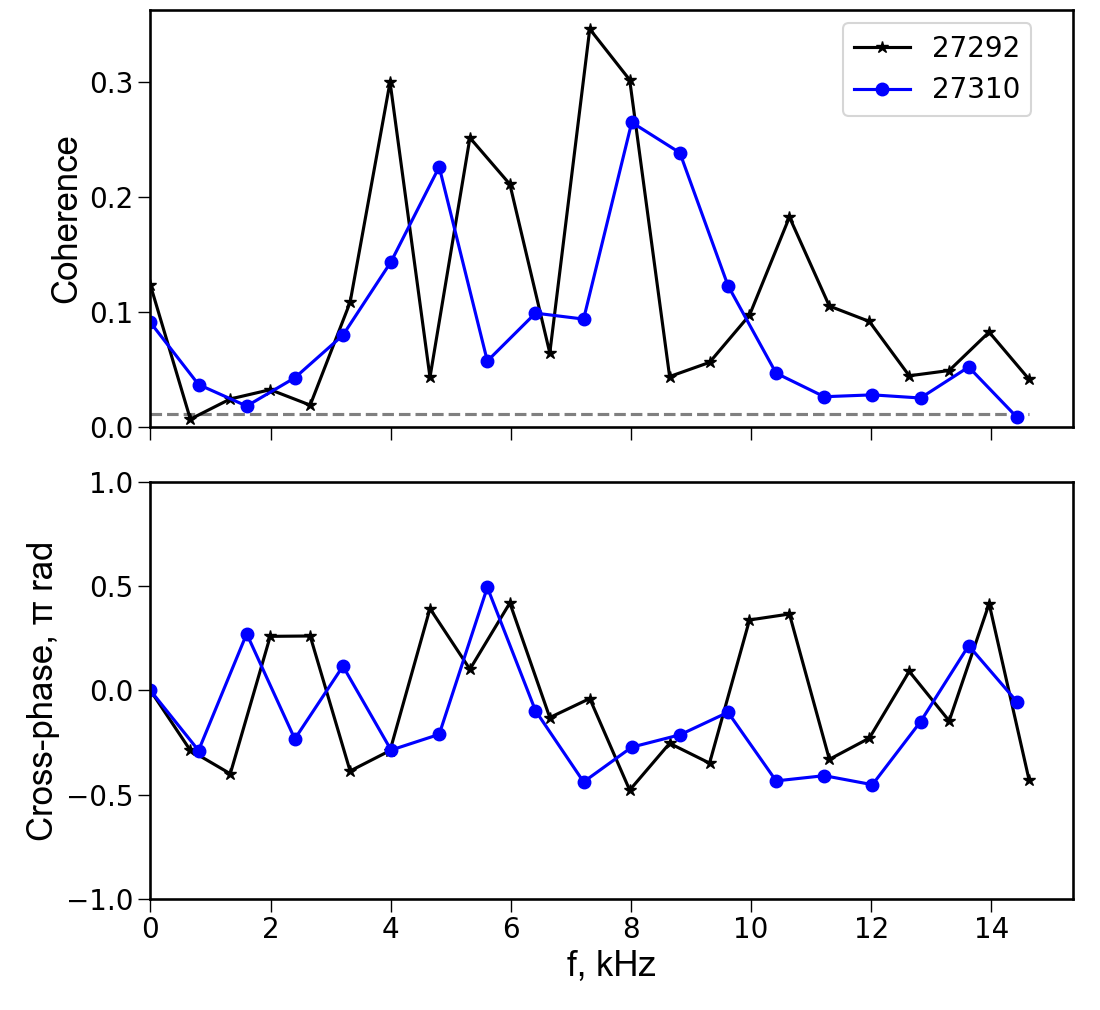}
    \caption{Cross-coherence and cross-phase spectra of poloidal velocity and density fluctuations for the shots $\#27292$, $t=0.110-0.120$ s, and \#$27310$, $t=0.113-0.125$ s, at $\Delta R=-7$ cm.}
	\label{fig: cohnv}

\end{figure}

We have found a correlation between the density and velocity fluctuations at 8 kHz across all viewing locations of the BES, negative cross-phases indicating that the density fluctuations lead velocity fluctuations were obtained only at $\Delta R=-7$ cm, $r/a=0.88$, between the blob$-$hole formation and hole-dominated regions (Figure \ref {fig: cohnv}). The cross-phase between $\delta n/n$ and $\delta v_{\perp}$ decreases towards the blob$-$hole formation region, where it is close to zero.

The broad oscillation peak at $r/a=0.92$ may be consistent with the co-existence of an 8 kHz GAM propagating from the core and excitation of modes at lower frequencies. A broad GAM has been previously measured at the edge of ohmic MAST plasma \citep{Hnat2018}. The GAM frequency of 8 kHz tends to persist over a range of observed radii. The electron temperature changes from 300 eV to 20 eV as $R$ varies from $R=1.12$ to 1.28 m; however, the GAM frequency, according to (\ref{w_GAM}), does not mirror these and other changes in the equilibrium parameters. This suggests a global or eigenmode character for the GAM structure. A global GAM structure has been observed previously both experimentally and in simulations \citep{Conway2022}. In particular, numerical simulations for MAST-relevant conditions have shown that the GAM frequencies form a set of the plateau (a set of radial regions with the constant GAM frequency), with several frequencies overlapping at one radial location \citep{Robinson2013}. This might explain the coherence spectra of $v_{\perp}$ in Figure \ref{fig: cohvv} that demonstrate multiple peaks at frequencies above 8 kHz. The peak splitting is also evident at frequencies below 8 kHz at $r/a=$ 0.97. Multiple peaks in spectra of perpendicular velocity oscillations were observed experimentally on various magnetic confinement devices \citep{Conway2008,Melnikov,Conway2022}.

High cross-coherence values and small cross-phases are observed in the lower-frequency part (below 4 kHz) of the velocity spectra: see Figure \ref{fig: cohvv}. This suggests the presence of a low-frequency branch of zonal flows. Although theory typically predicts zero-frequency zonal flows driven by turbulence \citep{Diamond2005, Ivanov2020}, in practice zonal flows have small finite frequencies. The study of nonlinear coupling between density fluctuations and LF zonal flows is presented in section \ref{nonlincoupl} and supports that these modes are turbulence-driven sheared flows.

\subsection{Shearing rates}
\label{Shear_rate}
It is widely accepted that zonal flows play a crucial role in regulating turbulence through $E \times B$ shearing of the eddies \citep{Rogers2000, Diamond2005, Ricci2006}. To assess the impact of the turbulence-driven flows, their spatio-temporal characteristics should be considered. The time scales on which zonal flows can stretch the density structures in the perpendicular direction and reduce their radial extent are compared here with the correlation times of density fluctuations.

The shearing rate was estimated as $\omega_{sh,GAM}=v_{rms,GAM}/l_{cor}$, where $v_{rms,GAM}$ is the rms amplitude of the perpendicular velocity at the GAM frequency ($v_{GAM}$), and $l_{cor}$ is the radial correlation length of the mode. 

The oscillatory velocity $v_{GAM}$ was obtained by bandpass filtering the total perpendicular velocity $v_{\perp}$ measured by the CCTDE method in the range $7-8.5$ kHz. The resulting rms amplitudes of the GAM were $v_{rms,GAM}=0.3-1$ km/s, while the mean velocity averaged over time was $\langle v_{\perp} \rangle=5-10$ km/s. Thus, the GAM rms amplitude constitutes $2-20\%$ of the mean perpendicular velocity.

\begin{figure}
    \centering
       
   \includegraphics[width=0.7\textwidth]{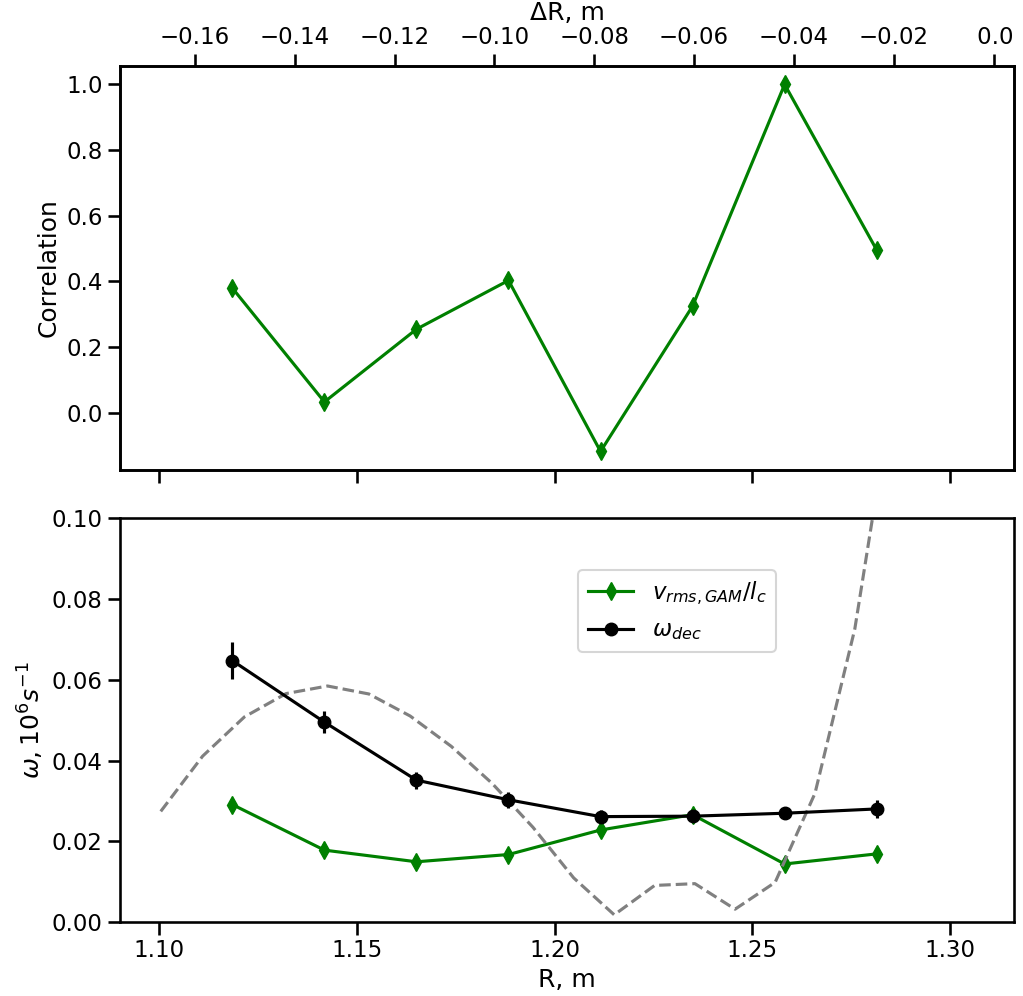}

    \caption{Top: Pearson cross-correlation coefficient between the rms of perpendicular velocities at the GAM frequency at different radial locations (reference location $R=1.26$ m). Bottom: turbulence decorrelation rate (circles) and the GAM-velocity shearing rate (diamonds) against major radius. The grey dashed line is the equilibrium perpendicular flow shear calculated using CXRS measurements. The data is for the shot $\#27292$, $t=0.115-0.119$ ms.}
	
	\label{fig: cc_vperp}
\end{figure}

To determine $l_{cor}$, we calculated the Pearson cross-correlation coefficient between the coherent flows at the GAM frequency at different radial locations, shown in Figure \ref{fig: cc_vperp}. One can see that there is a strong correlation between the poloidal velocities at the neighbouring radial locations. The radial correlation length of the GAM was estimated by fitting the cross-correlation coefficients by the function $e^{-\Delta r/l_{cor}}$, where $\Delta r$ is the distance to the reference channel. The correlation lengths were in the range of $2-4$ cm.

The time-averaged shearing rates of the GAM were approximately $20-30$ kHz, which is close to, and below, the density fluctuations' decorrelation rate of $30-60$ kHz (see Figure \ref{fig: cc_vperp}). The decorrelation rate $\omega_{dec}$ is defined as the inverse of the correlation time, calculated using the same technique as described in section \ref{cor_time_len}. These decorrelation rates are for the total density-fluctuation field, which includes both turbulent fluctuations and coherent structures, and turbulent fluctuations have shorter correlation times than coherent structures. In section \ref{v_vs_time}, it was shown that the GAM had intermittent behaviour, with a transient increase of GAM power by a factor of $2-8$ (Figure \ref{fig: wavelet_power}). Therefore, the GAM shearing rates also exhibited a transient increase during bursts in $v_{GAM}$. Thus, the shearing times are comparable to turbulence decorrelation times in between the bursts of the density-fluctuation power occurring due to either propagation of density holes or upshifts in the turbulence power. This is valid for the radial region $R=1.20-1.25$ m, $\Delta R=-4$ to $-9$ cm, where the mean $\omega_{sh,GAM}$ are close to $\omega_{dec}$ of the total density-fluctuation field. That region is where the density holes propagate radially from their formation region at $R=1.25$ m, $\Delta R=-4$ cm towards the core. Note that the shearing rates of the LF zonal flow defined analogously, were found to be close to the shearing rates of the GAM. The time evolution of the LF zonal flow also demonstrated a quasi-periodic increase in the amplitude of the flow velocity, so the shearing rates of the LF zonal flow were sufficiently high to have an impact on the density fluctuations.

\begin{figure}
    \centering
       
   \includegraphics[width=0.6\textwidth]{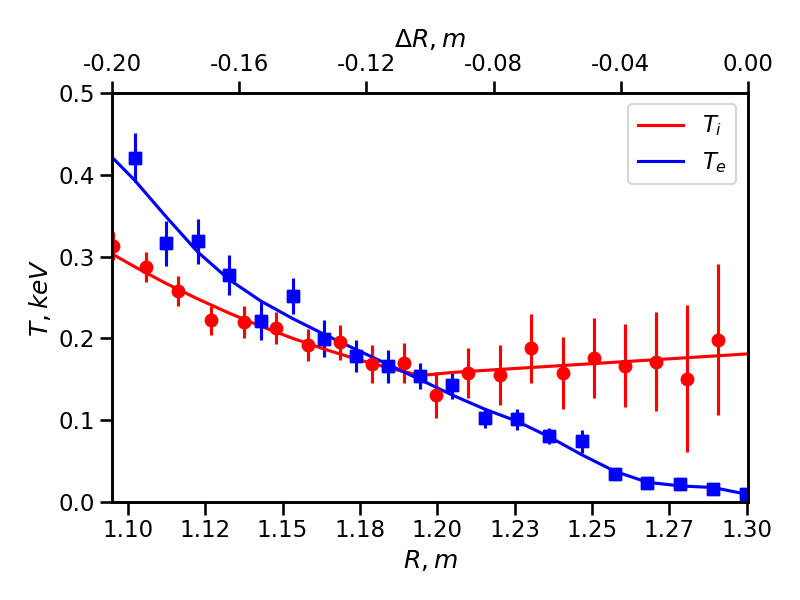}

    \caption{Electron (blue) and ion (red) temperature profiles in the shot $\#27292$ averaged over $t=0.115-0.119$ ms. Circles correspond to the CXRS measurements of the ion temperature. Squares correspond to the Thomson scattering measurements of the electron temperature.}
	
	\label{fig: te_ti_27292}
\end{figure}

 The background $E \times B$ flow shear, estimated using the CXRS flow-velocity measurements, was below the density-fluctuation decorrelation rate and the GAM shearing rate in the region dominated by the density holes: see the dashed line in Figure \ref{fig: cc_vperp}. Aside from the toroidal velocity, determining the total background $E \times B$ shearing rate also requires calculating the pressure gradient and the poloidal velocity. However, no measurements of the main ion density and poloidal velocity were available; therefore it was not possible to calculate the total mean $E \times B$ shearing rate. Figure \ref{fig: te_ti_27292} shows dissimilar behaviour of the ion and electron temperature profiles at the plasma edge, with the ion temperature profile flattening at $R=1.20-1.30$ m, $\Delta R=-10$ to $-7$ cm. This suggests that the ion-temperature-gradient component of the mean $E \times B$ flow should not be significant. Further work is required to evaluate the total background $E \times B$ shearing rate.
 
 As was shown in section \ref{Tilt_an}, in the case that we considered, the density-fluctuation field had almost zero tilt angle during the hole propagation and a finite tilt in between (Figure \ref{fig: tilt_angle}). A general trend of reduced tilting of the total density-fluctuation field in the region dominated by density holes compared to other plasma locations can be seen from Figures \ref{fig: sk_etc}, \ref{fig: cor_th_vr_27292_t_112}, and \ref{fig: vr_27292_27310}. If the mean $E \times B$ flow shear were sufficient to be responsible for tilting the background fluctuations, then it should have also sheared the density holes since they had longer lifetimes compared with the background fluctuations (Figure \ref{fig: ct}). This and the anticorrelation in time (or correlation with a time delay) between $v_{GAM}$ and $\delta n/n$ (Figure \ref{fig: wavelet_power}) suggest that it is the GAM that is responsible for shearing the background turbulent fluctuations. This does not rule out the possibility of scenarios where equilibrium flow shear plays a crucial role in regulating the dynamics of turbulent fluctuations.
 
 At other radial locations towards the plasma core, $R=1.11 - 1.19$ m, and towards the edge at $R=1.26-1.28$ m, $\omega_{sh,GAM}$ was somewhat below $\omega_{dec}$, although they were of the same order. Even though the GAM can transiently shear density fluctuations at those locations, it is hard to distinguish its impact from the equilibrium flow shear that is close to $\omega_{dec}$. It is within the region dominated by density holes that the GAM affects the tilt angle and radial correlation length of the ambient turbulent structures. 

\subsection{Nonlinear coupling}
\label{nonlincoupl}
Zonal flows are linearly stable plasma modes driven by nonlinear interactions. They are generated by Reynolds stresses associated with the unstable perturbations \citep{Diamond2005}. The effect of the Reynolds stress can be assessed using the normalised quantity $\langle \tilde{v}_p \tilde{v}_r \rangle / (\langle \tilde{v}_p \rangle \langle \tilde{v}_r \rangle)$, where $\tilde{v}_p=v_p - \langle v_p \rangle$ is the fluctuating poloidal velocity and $\tilde{v}_r$ is the fluctuating radial velocity, defined analogously. A statistically significant correlation between the fluctuating poloidal and radial velocities in the shot $\#27292$, $t=0.109-0.119$ ms, is observed near the blob$-$hole formation region at $R=1.26$ m ($\Delta R=-0.04$ m), characterised by strong inward motion. As we recall from section \ref{mode}, around that location we observed a peak at the GAM frequency in the poloidal velocity spectra. The significance of Reynolds stresses was also confirmed in this way at $R=1.14-1.18$ m ($\Delta R=-0.16$ to $-0.11$ m).

 The nonlinear coupling between density fluctuations and zonal flows in the frequency domain can be analysed using bispectral analysis \citep{Kim1979, Ritz1989}. The bispectrum reveals the strength of phase coupling between various spectral components of the fluctuating signals. If we consider three independent waves, their phases will be randomly distributed and statistical averaging over a sufficiently large number of realisations, or epochs, will result in a zero bispectrum. In contrast, if the spectral components are nonlinearly coupled to each other, the total phase difference is not random and the resulting bispectrum will show the strength of coupling between Fourier components. A sufficiently large number of evaluations of the triple product of the Fourier transforms of the signals is necessary to eliminate the random phase mixing that can be present in the data. 

\begin{figure}
    \centering
    \begin{subfigure}{0.7\textwidth}
      \includegraphics[width=\textwidth]{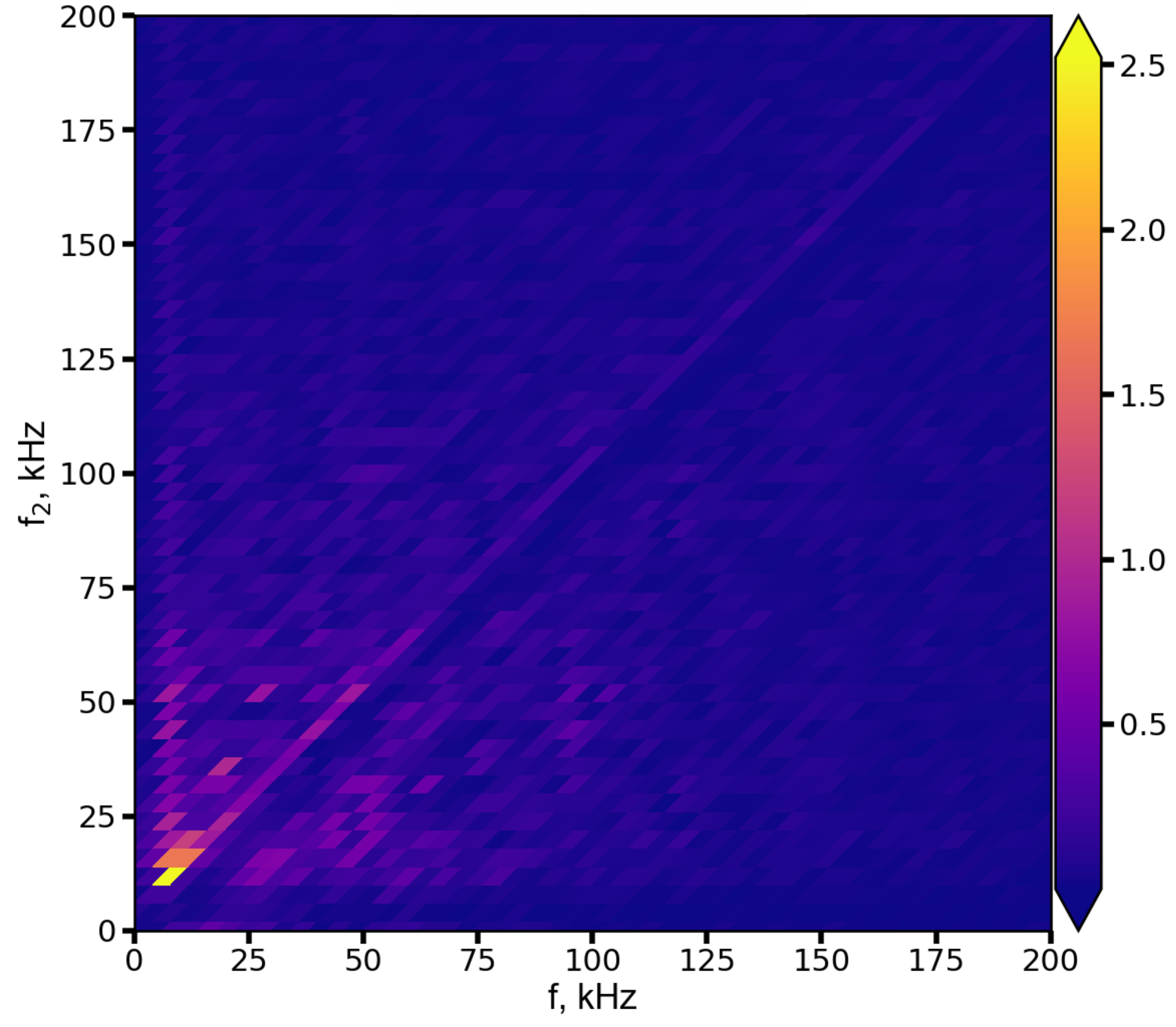}
      \caption{}
    \end{subfigure}
    \hspace{0.1cm}
    \begin{subfigure}{0.7\textwidth}
      \includegraphics[width=\textwidth]{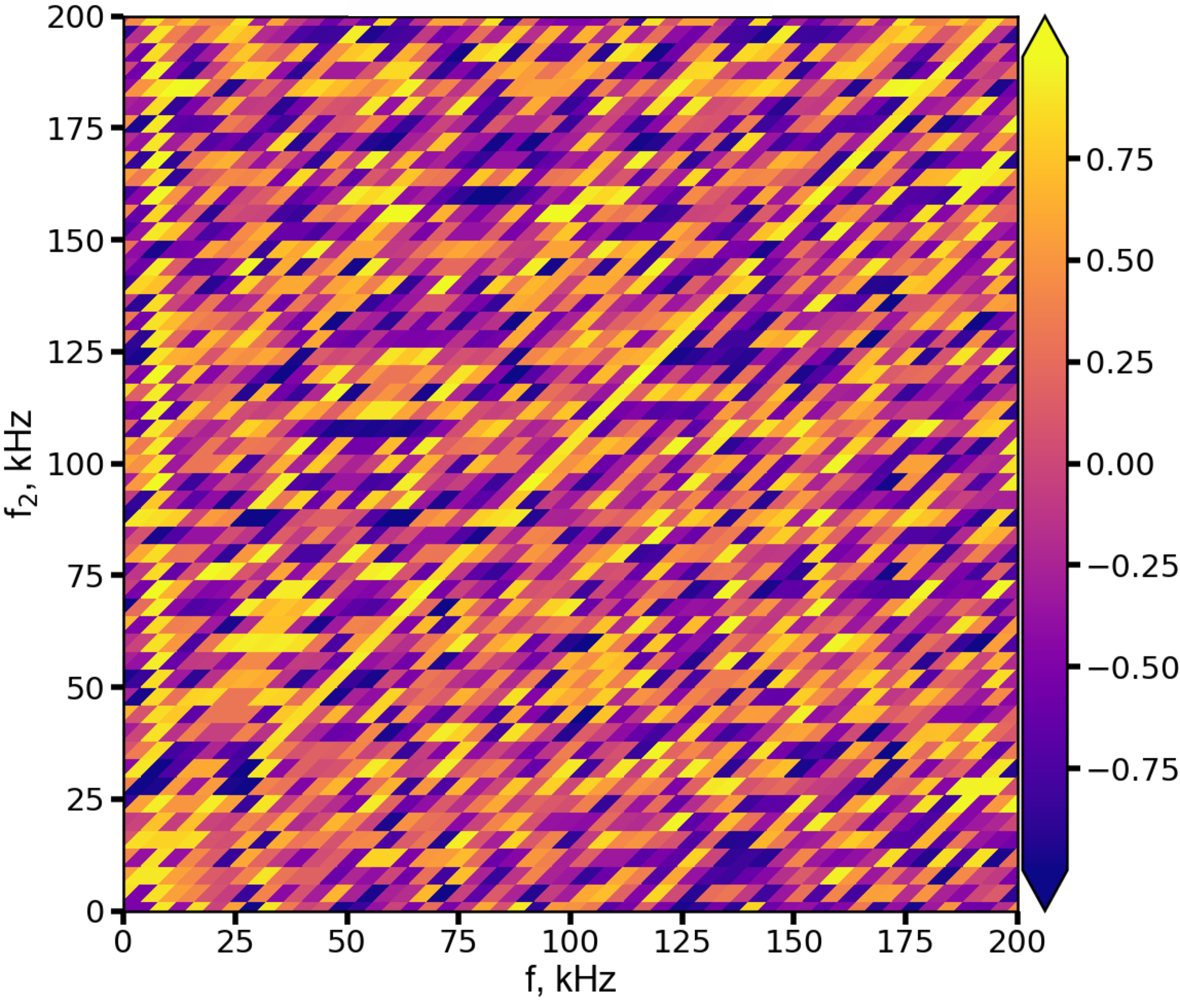}
      \caption{}
    \end{subfigure}
    \caption{(a) Absolute value of the autobispectrum $|\langle n_{f_1} n_{f_2} n_f \rangle|$ and (b) the biphase  of $\langle n_{f_1} n_{f_2} n_f \rangle$, $\pi$ rad, for the shot $\#$27292, $R=1.14$ m, $\Delta R=-0.16$ m. The frequency resolution of the spectra is 4 kHz.}
	\label{fig: bispec}
\end{figure}

 Since the GAM has a pressure-oscillation sideband associated with it, it is possible to use the autobispectral estimate of the density fluctuations to illustrate the nonlinear behaviour of the zonal mode. To calculate $\langle n_{f_1} n_{f_2} n_{f} \rangle$, the time period of 8.9 ms duration was divided into 1.5 ms time windows for averaging. A distinct feature of the resulting autobispectrum is the nonlinear coupling of a mode at frequency $f=|f_1|+f_2=8$ kHz with broadband turbulence with the span in frequencies from the lowest resolvable up to $f_2=$ 200 kHz (see Figure \ref{fig: bispec}(a)). Thus, the generation of the GAM is a result of three-wave interaction $-$ two unstable modes at frequencies $f_1$, $f_2$, and one zonal mode at $8$ kHz. A wide range of frequencies of fluctuations coupling to a zonal mode indicates cascading of energy to higher wavenumbers.

The triad interactions involving the GAM are expressed in the constant biphase, with a biphase value close to $\pi$ rad, whereas elsewhere biphases have random values ranging from $-\pi$ to $\pi$ (see Figure \ref{fig: bispec}(b)). The constant biphase indicates that all density-fluctuation modes couple to the same coherent potential fluctuation originating from a GAM, while interactions between density-fluctuation modes at other frequencies can have a wide range of biphases due to a variety of spatial structures \citep{Itoh2005}. Similar behaviour of the triple products and biphases was observed at $R=1.21$, 1.23, and 1.25 m, with the frequencies of unstable modes participating in the nonlinear interactions with the zonal flow spanning up to $70-100$ kHz. An oscillatory flow at 8 kHz was observed across $R=1.14 - 1.25$ m, suggesting that the GAM is a global phenomenon. These results are consistent with observations of autobispectra and biphases at GAM frequencies on other plasma devices \citep{Nagashima2006, Liu2010, Melnikov2017}.
 
We have so far considered triple correlations of a single scalar field $-$ the density-fluctuation field. Interactions between it and the sheared flows can also be studied using a two-field model for the density and velocity fluctuations that provides information about the direction of the energy transfer \citep{Holland2007,MXu2009,Cziegler2013}. Using this framework, we can work out an expression for the energy transfer\footnote{Note that by energy transfer, we mean the transfer of fluctuation power (or intensity) due to nonlinear coupling, without an implication that the variance of $\delta n$ is a conserved quantity.} term, a third-order cumulant in frequency space, characterising three-wave coupling. Let us start from the continuity equation and consider explicitly only the convective nonlinearity characterising the poloidal dynamics of the system:
\begin{equation}\label{eq_continue}
\partial_t \delta n +  v_y \partial_y \delta n = S,
\end{equation}
where $\delta n$ is the electron density fluctuation, $y$ is the poloidal coordinate, and $v_y$ is the poloidal velocity. The right-hand side of (\ref{eq_continue}) represents all other effects responsible for the changes in the density fluctuations such as the drive due to the equilibrium density gradient, compressibility, parallel dynamics, atomic processes, etc. A Fourier decomposition in frequencies is performed according to $\delta n = \sum_{f} \tilde{n}_f e^{i2\pi f t}$, whereby (\ref{eq_continue}) becomes:
\begin{equation}\label{eq_ftcont}
\partial_t \tilde{n}_f +  \sum_{f_1} \tilde{v}_{y,f-f_1} \partial_y \tilde{n}_{f_1}= \tilde{S}_f.
\end{equation}
Multiplying both sides of (\ref{eq_ftcont}) by $\tilde{n}_f^*$ and adding the result to its complex conjugate leads to an equation describing the evolution of the spectral power of the density fluctuations. We assume its time evolution to be slower than the interactions between modes. The resulting equation is
\begin{equation}\label{eq_tn}
\frac{1}{2}\partial_t \langle |\tilde{n}_f|^2 \rangle = -  \sum_{f_1} R e\langle \tilde{n}_f^*\tilde{v}_{y, f-f_1} \partial_y \tilde{n}_{f_1} \rangle + \gamma(f)  \langle |\tilde{n}_f|^2 \rangle,
\end{equation}
where $\langle ... \rangle$ denotes time averaging on scales longer than the interaction time scales but shorter than the time evolution of $\langle |\tilde{n}_f|^2 \rangle$. All linear driving and damping terms for density fluctuations are contained in the effective rate $\gamma $. The nonlinear coupling term 
\begin{equation}
\label{tn}
 T_n = - Re \langle \tilde{n}_f^*\tilde{v}_{y, f-f_1} \partial_y \tilde{n}_{f_1}  \rangle 
\end{equation}
represents energy transfer: interactions between fluctuations of the poloidal gradient of density at frequency $f_1$, the poloidal velocity fluctuations at frequency $f_2=f-f_1$ and the density fluctuations at frequency $f$ lead to energy transfer to (from) the density fluctuations. If $T_n(f_1,f)>0$, then the components at the target frequency $f$ gain energy from the density-fluctuation gradient at the source frequency $f_1$ or poloidal velocity fluctuations at frequency $f_2$. If $T_n(f_1,f)<0$ the components at the target frequency $f$ lose energy to the density-fluctuation gradient at frequency $f_1$ or poloidal velocity fluctuations at frequency $f_2$. 

The density fluctuation $\tilde{n}_f$ in (\ref{tn}) was calculated as an average between the fluctuations at two poloidal locations. The density-fluctuation gradient was calculated using the same locations as follows: $\partial_y \tilde{n} =(\tilde{n}_2-\tilde{n}_1)/\Delta y$, where $\tilde{n}_i$ is the relative density fluctuation at poloidal location $y_i$ and $\Delta y=y_2-y_1 $ is the separation between them. Note that the resulting $T_n$ is inherently antisymmetric with respect to the frequency $f_2$.

To calculate $T_n$, the time series of the shot $\#$27292, $t=0.109-0.119$ ms, were divided into intervals of 670 $\mu s$, and the averaging was performed over 12061 realisations, with the number of fully independent epochs of 11. The convergence of the calculations was confirmed via bispectrum calculations for different numbers of epochs.

\begin{figure}
    \centering
       
    \includegraphics[width=0.5\textwidth]{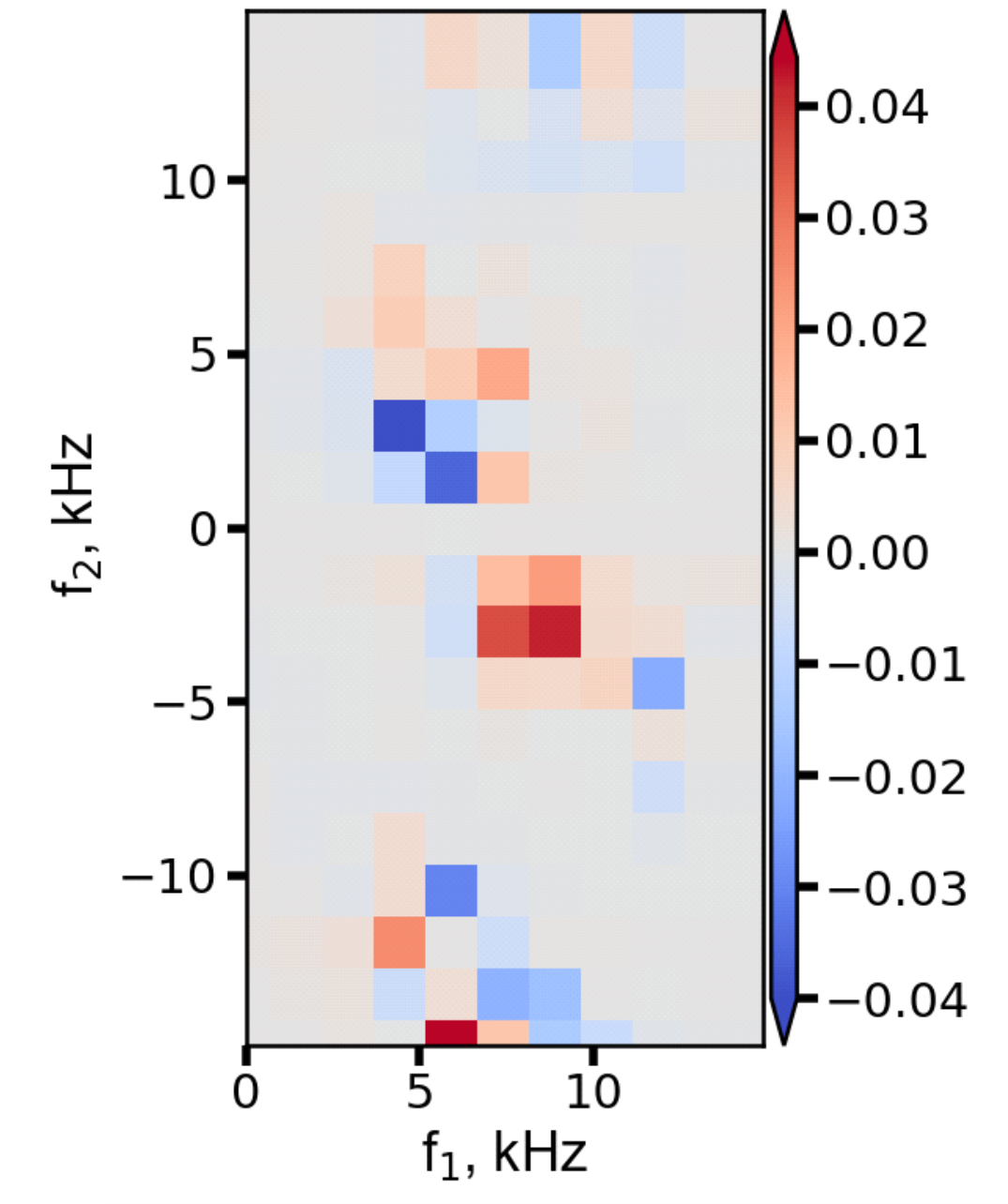} 

    \caption{Real part of the cross-bispectrum $T_n = - Re \langle \tilde{n}_f^*\tilde{v}_{y, f_2} \partial_y \tilde{n}_{f_1}\rangle$, a.u., for the shot $\#$27292, $R=1.23$ m, $\Delta R=-0.07$ m.}
	
	\label{fig: bispecnnn}
\end{figure}

Shown in Figure \ref{fig: bispecnnn}, the nonlinear term $T_n$ attests to a strong nonlinear coupling at $(f_1, f_2, f=f_1+f_2)=(9, -3, 6)$ kHz and $(4.5, 3, 7.5)$ kHz. According to (\ref{eq_tn}), the density fluctuations gain energy at 6 kHz and lose energy at 7.5 kHz. In both cases, this occurs through coupling to the poloidal-velocity fluctuations at $f_2=3$ kHz. The coherence spectra of poloidal velocities at two poloidal locations did show correlation above the noise floor and close to zero cross-phase at low frequencies within $0.6$ and $3.3$ kHz (Figure \ref{fig: cohvv}). A similar picture with the strong three-wave interaction at $f_2=1.5-3$ kHz was observed at adjacent radial locations. Thus, we conclude that the low-frequency poloidal flows are nonlinearly coupled to the density fluctuations. These observations reinforce the manifestation of these velocity-fluctuation modes as turbulence-driven sheared flows: the mode at $f_2=1.5-3$ kHz belongs to the low-frequency branch of zonal flows. 

The triple product $T_n$ allows us to determine if coupling exists at $f=f_1+f_2$; if it is negative, the zonal flows ($f_2$) $or$ density-fluctuation gradient ($f_1$) receive energy from the density fluctuations ($f$). Hence in our case, it is possible that the LF zonal flow receives energy from the higher-frequency density fluctuations and mediates energy transfer to lower-frequency density fluctuations. Other experimental findings suggest that LF zonal flows and GAM can both gain energy from higher-frequency density fluctuations (see \citet{Xu2010}, \citet{Conway2022} and references therein). A result somewhat similar to ours was observed by \citet{Lan2008}, where the triple product $T_n$ was constructed from model signals, with the signal corresponding to a GAM causing amplitude modulation of density fluctuations. This is in contradiction with results obtained by \citet{Holland2007} and \citet{Cziegler2013} using the same two-field model: there, energy was transferred to density fluctuations at higher frequencies in steps of a GAM frequency.

Limitations of the measurements and of the velocimetry algorithm resulted in a restricted frequency resolution and range of frequencies in the cross-bispectral analysis presented here. The signal-to-noise ratio and temporal resolution of the CCTDE technique should be improved to obtain a more comprehensive picture.

\section{Summary}
\label{summary}
In this paper, we focused on studying two intertwined aspects of the edge plasmas in the outboard mid-plane at $r/a=0.8-1.1$: ion-scale density fluctuations and zonal flows. The results of the study can be summarised in two parts. The main conclusions related to the properties of edge density fluctuations are given below.

\begin{enumerate}

 \item Density fluctuations in the edge of MAST plasmas exhibit intermittent behaviour that is reflected in the tails of their PDFs. Intermittency of the fluctuations across different radial locations is a result of the formation and transport of large-amplitude, large-scale structures $-$ blobs and density holes. 

 \item The formation region of blobs and holes, defined as a region of zero skewness, was observed at a distance to the separatrix of up to 5 cm, with the majority of observations at $2-4$ cm inside the LCFS. The skewness profile was found to be sensitive to the change in the plasma equilibrium profiles, with the zero-skewness region located close to the local maximum of the normalised electron pressure gradient.
 
 \item Symmetry breaking of the PDFs of the density fluctuations towards the core and the edge from the blob$-$hole formation region is a result of the radially inward and outward movement of the density holes and blobs, respectively. The observed quadratic dependence of the kurtosis on the skewness suggests that a common mechanism is responsible for setting the statistical properties of the density-fluctuation field across different radial locations spanning 20 cm.

 \item Strong inward radial velocities of up to $-8$ km/s were measured between the formation region and the region dominated by the density holes. The density holes propagate up to 10 cm inside the LCFS where they eventually decorrelate. We observed that long-lived structures have longer correlation times, smaller tilt angles, and higher inward velocities than ambient turbulence.

\end{enumerate}

Using the velocimetry technique, the fast dynamics of poloidal flows and their interplay with density fluctuations were analysed. The results of this analysis can be summarised as follows:
\begin{enumerate}

 \item Poloidal flows exhibit bursty behaviour, with rises in the poloidal velocity alternating with increases in the density-fluctuation power. Such alternating dynamics of GAM and density fluctuations were evident at locations close to the blob$-$hole formation region. These findings suggest that not only generic turbulent fluctuations but also coherent structures (density holes) can cause the generation of zonal flows. These poloidal flows in turn regulate turbulence by suppressing the underlying fluctuations. The dynamics of the density-fluctuation intensity and zonal flows presented in this work are similar to the picture observed in the ITG fluid simulations of \citet{Ivanov2020, Ivanov2022}, where sudden increases in the heat flux due to coherent structures were followed by the growth in the amplitude of sheared zero-frequency flows.
  
 \item We have shown that LF zonal flows and the GAM are sustained sufficiently long and have shearing rates comparable to fluctuation decorrelation rates, so they can impact the ambient turbulence through shearing. This is also supported by the observation of larger tilt angles of the turbulent density fluctuations compared to the density holes. Density holes exhibit longer correlation times than background turbulence. Therefore, if the equilibrium flow shear were responsible for shearing background turbulence, it should have also sheared the density holes. However, this does not necessarily rule out the possibility of scenarios in which equilibrium flow shear plays a crucial role in regulating the dynamics of turbulent fluctuations. 

 \item Measurements of autobispectrum of density fluctuations revealed that the GAM was driven by broadband turbulence with frequencies up to 200 kHz. Such a wide range of frequencies of fluctuations coupling to the GAM supports that the mode transfers energy to higher wavenumbers. This is also suggested by the observation of higher tilt angles of the density structures when the zonal mode activity was maximal.
 
 \item Using cross-bispectral estimates, we have found that the low-frequency branch of zonal flows nonlinearly couples to the density fluctuations and mediates the transfer of density-fluctuation power. The sign of the transfer term indicated the transfer of the density-fluctuation power from high to low frequencies. This confirms that the observed velocity oscillations are LF zonal flows. However, further work is required to quantify the relative importance of LF zonal flows in comparison with GAM in regulating the dynamics and properties of density fluctuations.
 
\end{enumerate}

\section{Conclusions}
\label{conc}
Intermittent behaviour of density fluctuations associated with the presence of blobs and holes was observed at the edge of L-mode plasmas in MAST. By using statistical analysis it was found that coherent structures are born near or inside the separatrix at $\Delta R=-5-0$ cm. The blob$-$hole birth zone is localised close to the maximum of the normalised pressure gradient supporting an interchange mechanism of structure formation. Blobs and holes acquire radial velocity that allows the density holes to penetrate up to $\Delta R = -10$ to $-15$ cm ($r/a=0.83-0.86$). The results are qualitatively consistent with predictions of existing models \citep{Krasheninnikov2008,DIppolito2011} for the amplitude and direction of the radial velocities of coherent structures and present evidence of the impact of density holes on the fluctuation evolution.

We have found that intermittency of perpendicular-flow oscillations is governed by the dynamics suggestive of a predator-prey behaviour: an increase in the density-fluctuation power originating due to both broadband turbulence and propagation of density holes is followed by a rapid rise of perpendicular velocity fluctuations. This suggests that coherent structures are part of a quasi-periodic self-regulating system that exhibits a behaviour similar to the standard framework of the drift wave-zonal flow turbulence \citep{Rogers2000, Diamond2005, Kobayashi2012, Ivanov2020, Ivanov2022}: the kinetic energy is transferred from the fluctuations to sheared flows that suppress the initial drive and then decay allowing the perturbations to grow. The implications for magnetic confinement studies are that coherent structures, in particular, density holes, can impact the evolution of plasma turbulence by triggering amplification of zonal flows and are hence crucial in understanding the mechanisms of regulation of edge plasma transport not only in the SOL but deep into the confined region.

This study applies to the peripheral region of L-mode plasmas, where there is significant ion-scale turbulence. Although the formation of blobs has been observed in the SOL of H-mode plasmas on MAST \citep{Ayed2009}, further studies are required to determine whether similar dynamics are present in improved-confinement regimes. Since coherent structures are part of a self-regulating system, one can conjecture their potentially non-negligible role in the plasma dynamics leading towards a transition to an H-mode. Future work should be focused on understanding the impact of density holes on transport in scenarios relevant to future fusion power plants.

\section*{Funding}

This work (A.S., I.C., A.A.S. and P.G.I.) was supported by the EPSRC Programme grant EP/R034737/1. The work of A.A.S. was also supported in part by the STFC grant ST/W000903/1 and by the Simons Foundation via a Simons Investigation Award.

\section*{Declaration of Interests}

The authors report no conflict of interest.

\bibliographystyle{jpp}

\bibliography{refs}

\end{document}